\documentclass[11pt]{article}
\usepackage[dvipsnames]{xcolor}
\usepackage{CJKutf8}

\usepackage{amsthm,amsmath,color,natbib}
\usepackage{setspace}
\usepackage{graphicx}
\usepackage[bottom]{footmisc}

\newtheorem{Theorem}{Theorem}

\newtheorem{Lemma}{Lemma}
\newtheorem{Proposition}{Proposition}
\newtheorem{Corollary}{Corollary}

\newcommand{\p}{{\rm I}\kern-0.18em{\rm P}}
\newcommand{\1}{{\rm 1}\kern-0.24em{\rm I}}
\newcommand{\E}{{\rm I}\kern-0.18em{\rm E}}

\newcommand{\eqdef}{\mathrel{\mathop=}:}
\DeclareMathOperator*{\argmin}{arg\,min}

\usepackage{amsthm,amsmath,amsfonts,amssymb,natbib,mathtools,mathrsfs,algorithm,framed,multirow}
\usepackage[noend]{algpseudocode}
\usepackage{ccaption}
\usepackage{longtable}

\usepackage{enumerate}
\usepackage{verbatim}
\usepackage{subfigure}
\usepackage{bbm}
\usepackage{threeparttable}

\usepackage{xr}

\usepackage{JASA_manu}
\usepackage{authblk}

\usepackage[colorlinks,linkcolor=blue,citecolor=blue,urlcolor=blue,bookmarksopen=true]{hyperref}

\makeatother

\begin{document}

\title{Intentional Control of Type I Error over Unconscious Data Distortion: a Neyman-Pearson Approach to Text Classification}

\author[1]{Lucy Xia}
\author[2]{Richard Zhao}
\author[3,5]{Yanhui Wu}
\author[4,5]{Xin Tong}
\affil[1]{\footnotesize Department of ISOM, School of Business and Management, Hong Kong University of Science and Technology.}
\affil[2]{\footnotesize Department of Computer Science and Software Engineering, The Behrend College, The Pennsylvania State University. }
\affil[3]{\footnotesize Faculty of Business and Economics, University of Hong Kong; Department of Economics and Finance, University of Southern California. }
\affil[4]{\footnotesize Department of Data Sciences and Operations, Marshall School of Business, University of Southern California. }
\affil[5]{To whom correspondence should be addressed. yanhuiwu@marshall.usc.edu, xint@marshall.usc.edu}

\date{}

\maketitle

\begin{abstract}

This paper addresses the challenges in classifying textual data obtained from open online platforms, which are vulnerable to distortion. Most existing classification methods minimize the overall classification error and may yield an undesirably large type I error (relevant textual messages are classified as irrelevant), particularly when available data exhibit an asymmetry between relevant and irrelevant information. Data distortion exacerbates this situation and often leads to fallacious prediction. To deal with inestimable data distortion, we propose the use of the Neyman-Pearson (NP) classification paradigm, which minimizes type II error under a user-specified type I error constraint. Theoretically, we show that the NP oracle is unaffected by data distortion when the class conditional distributions remain the same. Empirically, we study a case of classifying posts about worker strikes obtained from a leading Chinese microblogging platform, which are frequently prone to extensive, unpredictable and inestimable censorship. We demonstrate that, even though the training and test data are susceptible to different distortion and therefore potentially follow different distributions, our proposed NP methods control the type I error on test data at the targeted level. The methods and implementation pipeline proposed in our case study are applicable to many other problems involving data distortion. 	

\vspace*{.15in}
	
\end{abstract}

\section{Introduction}

The rise of social media platforms has spurred the extensive use of large-scale textual data for both academic and non-academic purposes. However, textual data on open digital platforms are susceptible to manipulation, evident from the continuous debates about fake news, censorship, internet trolls, and social bots \citep{Woolley.Howard.2016a,Woolley.Howard.2016b}. Within an environment of data distortion, the utilization of textual data for information collection (e.g., gauging public opinion) and event discovery (e.g., monitoring social unrest) can be challenging. In the context of textual classification, this paper shows the powerlessness of existing classification approaches to handling unknown or inestimable data distortion. We then propose and illustrate the use of the recently developed Neyman-Pearson (NP) classification approach that aims to asymmetrically control classification errors \citep{CanHowHus02, Sco05, Rigollet.Tong.2011, Li.Tong.2016, tong2016np} in some common situations of data distortion, such as data obtained from censored Chinese social media.

Since 2009 when \textit{Sina Weibo} – the Chinese equivalent to Twitter – was launched, social media have created an unprecedented informational shock to the Chinese society. Notably, Sina Weibo enables millions of citizens to generate and communicate political information that is scarce in traditional media. Government agents, media outlets, NGOs and firms, and researchers have invested heavily in machine learning techniques to mine the wealth of textual information circulated on Sina Weibo \citep{Economist.2013, ChinaInternet.2013, ChinaInternet.2014}. However, due to the potential effect of widespread political information on social unrest and regime stability, the Chinese government extensively censors social media \citep{chen2011internet, king2013censorship, king2014reverse}. Such censorship gives rise to two major challenges faced by data analysts in their endeavor of text mining. First, although the Chinese government allows for relatively free information flow on social media for the purposes of surveillance and monitoring officials \citep{qin2017does}, censorship substantially reduces the amount of information circulating on social media that can practically be used to classify data and predict hidden social events. The objective of minimizing the overall classification error, which is used by most existing machine learning algorithms, can cause an undesirably large error of missing important information. Second, social media censorship in China relies mostly on ad hoc human manipulation to fine-tune the extent of censorship in response to the changing local and temporal social conditions (\citep{Bamman2012censorship, Zhu2013velocity}). This censorship strategy makes it infeasible to infer the censorship rate. Thus, the traditional solution that corrects the potential bias due to data truncation through a parametric estimation of the censorship rate is hardly a practical choice. We propose the use of the NP classification approach to precisely overcome these two challenges.

To make our discussion more concrete, consider that a decision maker wishes to use social media posts about political issues and social events to discover and monitor grass-root political actions such as protests, petitions, or worker strikes. To this end, the decision maker must use algorithms trained on labeled data to classify a large number of posts, i.e., to predict discrete outcomes (class labels) for upcoming posts. In a binary classification setting, a post is coded in $\{0, 1\}$, where class $0$ means relevant to a specific topic, and class $1$ means irrelevant. Two types of errors occur: type I error (mislabel class $0$ as class $1$) and type II error (mislabel class $1$ as class $0$).\footnote{In the verbal discussion, type I and type II errors can also be thought of as the probability of making such errors.} The default classification objective in practice, which is referred to as \textit{the classical classification paradigm} in this paper, is the one that minimizes the overall classification error, which is a weighted sum of type I and type II errors, with weights being the proportions of classes. When controlling one type of error is dominantly important, a conflict occurs between the need for asymmetric control over classification errors and the neglect of such consideration in the overall classification error. Data distortion can exacerbate such a conflict. If a fraction of class $0$ data is eliminated, then in the objective function of the classical paradigm, the weight of type I error is reduced. Minimizing this objective naturally increases type I error, which is undesirable when controlling type I error to avoid overlooking relevant events is crucial to decision making.

In this paper, we first derive the \textit{classical oracle classifier} (theoretically optimal classifier under the classical paradigm) regarding the post-distortion population, and then demonstrate that, without precise knowledge about the data distortion rates, the pre-distortion classical oracle classifier cannot be recovered even if we have access to the entire post-distortion population. As a solution, we propose to use the Neyman-Pearson classification paradigm (NP paradigm) which minimizes type II error under a user-specified type I error constraint. The NP paradigm has the advantage that the NP oracle classifier (theoretically optimal classifier under the NP paradigm) is invariant to the class size proportion in the population. This property guarantees the invariance of the NP oracle under any distortion scheme as long as the class conditional distributions of the features remain the same.

To bring our theoretical discussion to live, we focus on an exemplary case in the general setting of Chinese social media, in which we classify a large number of posts about worker strikes published on Sina Weibo. Accurately identifying strike events in a timely manner is highly valuable for many decision makers, including governments, firms, and social scientists studying social movement. On the other hand, as a type of collective action, posts about strikes are prone to censorship, the extent of which varies across regions and over time. We show that applying existing classification methods leads to a considerable type I error, which can result in oversight or fallacious outcomes in decision making. We then use an NP umbrella algorithm \citep{tong2016np} in combination with state-of-the-art machine learning techniques to classify the posts. Consistent with the NP oracle's invariance property to data distortion, we find that even though the training and test data are susceptible to different distortion rates and are thus differently distributed, the NP classifiers hold type I errors well controlled at the targeted level on the test data. Furthermore, we demonstrate that for the purpose of controlling type I errors, the NP classification methods allow decision makers to borrow data generated in an information-abundant environment to classify data generated in an information-scarce environment. This advantage is important when decision making is constrained by time and resources.

Our study of data distortion is essentially an inquiry into the validity of statistical prediction when the process of data generation is a primary concern. This concern is not dismissible even in the era of big data. Instead, it can be exacerbated when data sources are vulnerable to human intervention. One candidate solution to data distortion is to estimate and correct potential bias by assuming precise knowledge regarding data generation and distortion. This is analogous to the parametric estimation of censored or truncated data in classical statistical inference \citep{Chung.Schmidt.Witte.Witte.1991}. Unfortunately, such a solution is infeasible when data are generated from diverse sources and are affected by complex interactions. Another potential solution, which is popular in the traditional statistics literature, is the development of sampling techniques that aim to obtain more representative samples from the population \citep{Luborsky.Rubinstein.1995}. However, sampling methods do not solve the data distortion problem in our study because even if the entire post-distortion population were available, knowledge about the pre-distortion population is still limited by unknown or inestimable distortion rates. In contrast, the NP classification approach we propose allows researchers to bypass one common kind of distortion which changes the class proportions but not the class conditional feature distributions.

The setting in this paper might seem similar to domain adaptation \citep{BenDavid2010,Chen2011}, a type of transfer learning.  However, the data distortion problem in our study differs fundamentally from the problems studied in domain adaptation. In domain adaptation, a key assumption is that the ``source domain" and ``target domain" share the same feature space, but have different feature distributions. A domain adaptation algorithm takes not only labeled data from the source domain, but also data (labeled or unlabeled) from the target domain. In contrast, the only available training data in our study are the labeled data from the post-distortion population (i.e., the source domain) without using any data (regardless of being labeled or unlabeled) from the pre-distortion population (i.e., the target domain). In this sense, the data-distortion problem we address is more challenging because data from the target domain is not available. To overcome such a challenge, the NP classification approach invokes the assumption that the features have the same conditional distributions in the source and target domains.

\section{Classification and Unknown Distortion Scheme}\label{sec:unknown distortion}

Binary classification is a supervised learning procedure frequently used in textual analysis. It aims to classify a piece of textual message into a category that is relevant to either a specific purpose or an irrelevant category. Formally, the aim of binary classification is to accurately predict class  labels (i.e., $Y=0$ or $1$) for new observations (i.e., features $X\in\mathbb{R}^d$) on the basis of labeled training data. \textcolor{black}{For the rest of the discussion, we treat the relevant information category as class $0$ and the irrelevant one as class $1$, so that missing a class $0$ message is more consequential than missing a class $1$ message.}    %
Concretely,  let $h: \mathbb{R}^d\rightarrow \{0,1\}$ be a binary classifier, $R_0(h):= \p(h(X)\neq Y| Y=0)$ denote type I error, and $R_1(h):= \p(h(X)\neq Y|Y=1)$ denote type II error. Then, the (population) classification error $R(h)$ can be decomposed as
$
R(h)=R_0(h)\cdot\p(Y=0) + R_1(h)\cdot\p(Y=1)\,.
$
We use the term \textit{classical paradigm} to refer to the learning objective of minimizing $R(\cdot)$. The classical oracle classifier, i.e., the classifier that minimizes $R(\cdot)$ among all functions, is $h^*(x)  =  \1(\eta(x) > 1/2)$, where $\eta(x) = \E(Y |X=x)= \p(Y=1|X=x)$. The classical oracle $h^*$ is achievable only if the entire population is available. In practice, we have to train a classifier based on a  finite sample. %

\subsection{\textcolor{black}{Data Distortion Scheme}}
In this paper, we restrict our attention primarily to the type of distortion that changes the class proportion of the population without changing the class conditional distributions of the features. In other words, we assume that distortion changes $\p(Y=0)$ and $\p(Y=1)$, but does not change the distributions of $X|(Y=0)$ or $X|(Y=1)$.  The assumption that features have the same class-conditional distributions is justified if the distortion scheme in the dataset (e.g., deleting sensitive social media posts) is random. We will show that this assumption can be approximated by the data distortion situation in our case study and other real world applications. We discuss more general conditions in Appendix C.

\subsection{Oracle under Data Distortion}

\textcolor{black}{Denote the class $0$ distortion rate by $\beta_0 = \beta_0^- -\beta_0^+$,  where $\beta_0^-$ is  \textit{the class $0$ downward-distortion rate}  and  $\beta_0^+$ is  \textit{the class $0$  upward-distortion rate}. These rates are the proportions of class $0$ texts that are randomly deleted or injected, respectively. For example, $(\beta_0^-, \beta_0^+) = (.2, .1)$ means $20\%$ of class $0$ texts are randomly deleted from the population, and $10\%$ of class $0$ texts are artificially injected, so the net effect is a $\beta_0 = 10\% = 20\% - 10\% $ decrease in class $0$ texts.  Since we cannot disentangle the upward and downward forces just from the post-distortion population, we will formulate the theory only on the net decrease effect $\beta_0$.      Similarly, $\beta_1$ is defined for class $1$. Below, we derive the formula of the (classical) oracle classifier regarding the post-distortion population.}

\begin{Theorem}\label{prop: general classical oracle under distortion}
Let $f_0$ and $f_1$ denote the pre-distortion probability density functions  of  $X|(Y=0)$ and $X|(Y=1)$, and $\pi_0 = \p(Y=0)$ and $\pi_1 = \p(Y=1)$ be the class priors. Suppose the distortion scheme does not change the distributions for $X|(Y=0)$ and $X|(Y=1)$ but only the class proportions. Let $\beta_0$ and $\beta_1$ be the distortion rates of class $0$ and class $1$ respectively. Then, the classical oracle classifier regarding the pre-distortion population is %
\vspace{-0.05in}
$$h^*(x) = \1\left(\frac{f_1(x)}{f_0(x)} > \frac{\pi_0}{\pi_1}\right)\,,$$
\vspace{-0.05in}
and that regarding the post-distortion population is $$h^*_{(\beta_0, \beta_1)}(x) = \1\left(\frac{f_1(x)}{f_0(x)} > \frac{1-\beta_0}{1-\beta_1}\cdot\frac{\pi_0}{\pi_1}\right)\,.
$$ %

\end{Theorem}
\vspace{-0.03in}
In this theorem, the explicit analytic form of the classical pre-distortion oracle classifier $h^*$ is a well-known result, while that of $h^*_{(\beta_0, \beta_1)}$ is new. See Appendix A for its proof. The thresholds of $f_1/f_0$ in oracle classifiers $h^*$ (pre-distortion) and $h^*_{(\beta_0, \beta_1)}$ (post-distortion) differ by a multiplicative constant $(1-\beta_0)/(1-\beta_1)$. This difference in thresholds reflects a change in the class proportions in the population. If the entire post-distortion population is available, we can calculate the class conditional densities $f_0$ and $f_1$ as well as the post-distortion class proportions
$$
\pi^{(\beta_0, \beta_1)}_{0} = \frac{(1-\beta_0) \pi_0}{(1-\beta_0) \pi_0 + (1-\beta_1)\pi_1}\,, \text{ }
\pi^{(\beta_0, \beta_1)}_{1} = \frac{(1-\beta_1) \pi_1}{(1-\beta_0) \pi_0 + (1-\beta_1)\pi_1}\,.
$$
 Then, $h^*_{(\beta_0, \beta_1)}$ can be recovered. However, there is no hope to recover or estimate $h^*$, unless $\beta_0$ and $\beta_1$ are known or estimable.

\subsection{Impact of Censorship Rate under the Gaussian Model}
To visualize and quantify the result in Theorem 1, we study an example with $\beta_0> 0$ and $\beta_1 = 0$  under a canonical linear discriminant analysis model.
Let $f_0\sim \mathcal{N}(\mu_0, \Sigma)$ and $f_1\sim \mathcal{N}(\mu_1,\Sigma)$,  %
 where $\mu_0$ and $\mu_1$ represent mean vectors for classes $0$ and $1$ respectively, and $\Sigma$ is the common covariance matrix.
In this model, the decision boundary of the oracle $h^*$ is
\begin{equation}\label{eqn:oracle_decision_boundary}
\footnotesize{
x^{\top} \Sigma^{-1} (\mu_0-\mu_1) -\frac{1}{2} (\mu_0-\mu_1)^{\top}\Sigma^{-1}(\mu_0+\mu_1) + \log\left(\frac{\pi_0}{\pi_1}\right) =0\,.}
\end{equation}
\vspace{-0.05in}
When only $(1-\beta_0)$ proportion of observations from class $0$ remains, the post-distortion oracle classifier $h^*_{\beta_0}:=h^*_{(\beta_0, 0)}$
 has the following decision boundary:
\begin{equation}\label{eqn:oracle_decision_boundary_censored}
\footnotesize{
x^{\top} \Sigma^{-1} (\mu_0\hspace{-0.05cm}-\hspace{-0.05cm}\mu_1) \hspace{-0.05cm}-\hspace{-0.05cm}\frac{1}{2} (\mu_0\hspace{-0.05cm}-\hspace{-0.05cm}\mu_1)^{\top}\Sigma^{-1}(\mu_0\hspace{-0.05cm}+\hspace{-0.05cm}\mu_1) \hspace{-0.05cm}+\hspace{-0.05cm} \log\hspace{-0.05cm}\left(\hspace{-0.1cm}\frac{(1\hspace{-0.05cm}-\hspace{-0.05cm}\beta_0) \pi_0}{\pi_1}\hspace{-0.1cm}\right) \hspace{-0.1cm}=\hspace{-0.05cm}0\,.}
\end{equation}

\begin{figure}
\includegraphics[scale=0.27]{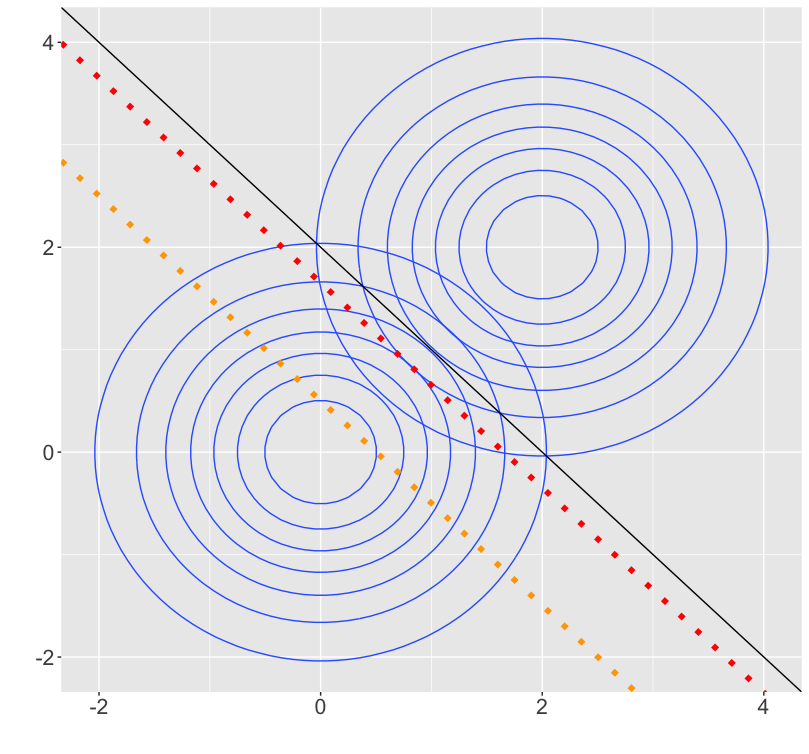}
\includegraphics[scale=0.31]{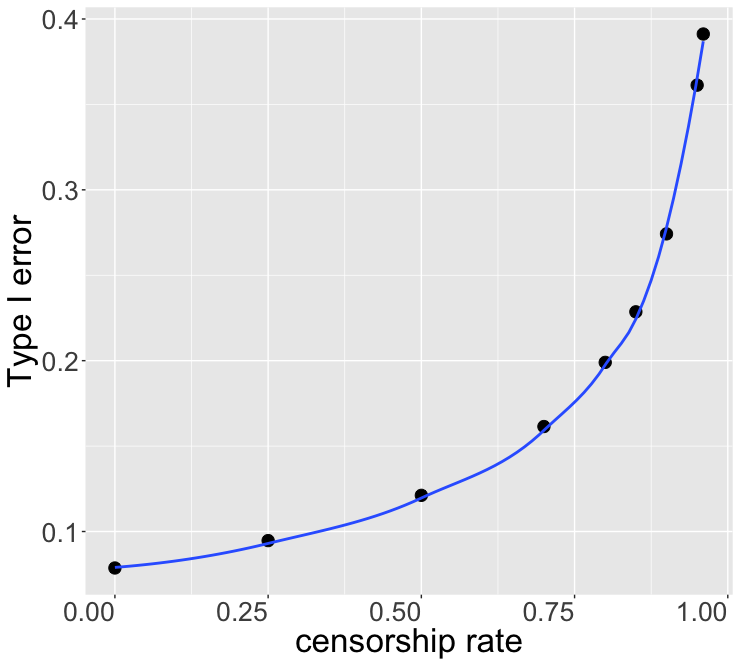}
\caption{\footnotesize{The left panel shows the shift of the oracle decision boundary due to distortion under a linear discriminant analysis model: $\mu_0 = (0,0)^{\top}$, $\mu_1 = (2,2)^{\top}$, $\Sigma=I$, $\pi_0=.5$. The horizontal axis and vertical axis are the two feature measurements, and the contours represent different density levels of each class.      The black line is the original oracle decision boundary; the red dashed line and the orange dashed line are the oracle decision boundaries after censorship on class $0$ with $\beta_0 = .5$ and $\beta_0 = .95$, respectively. The right panel plots type I error of $h^*_{\beta_0}$ as a function of $\beta_0$.  } \label{fig:censorhip_difference}}
\end{figure}

Comparing \eqref{eqn:oracle_decision_boundary} and \eqref{eqn:oracle_decision_boundary_censored}, the shape of the decision frontier remains the same, but the left hand of the equations differs by a constant   $\log(1-\beta_0)$. To visualize the difference in decision boundaries, we plot an example in  Figure \ref{fig:censorhip_difference}. Proposition \ref{prop:1} below further explores the relationship between type I error and the censorship rate of class $0$ for balanced classes.

\begin{Proposition}\label{prop:1}
Suppose the probability densities of class $0$ ($X|Y=0$) and  class $1$ ($X|Y=1$) follow distributions $\mathcal{N}(\mu_0,\Sigma)$ and $\mathcal{N}(\mu_1,\Sigma)$ respectively, and the two classes are balanced in the pre-distortion population (i.e., $\pi_0 = \pi_1 =.5$).  Suppose that the censorship rate of class $0$ is $\beta_0\in(0,1)$ and class $1$ is not distorted ($\beta_1 = 0$). To keep notations simple, let  $h^*_{\beta_0} = h^*_{(\beta_0,0)}$ be the classical oracle classifier in the post-distortion population. Then, the type I error of $h^*_{\beta_0}$ is:
\vspace{-0.05in}
\begin{equation}\label{type_i_error_censored}
R_0(h^*_{\beta_0}) = \Phi\left(\frac{-\frac{1}{2}C-\log\left(1-\beta_0\right)}{\sqrt{C}}\right)\,,
\end{equation}
where $C = (\mu_0 - \mu_1)^{\top}\Sigma^{-1}(\mu_0 - \mu_1)$.  Clearly, $R_0(h^*_{\beta_0})$ increases with $\beta_0\in (0, 1)$.
\end{Proposition}

Proposition \ref{prop:1} is proved in Appendix E. When censorship on class $0$ texts intensifies, class $0$ in the post-distortion population represents a smaller proportion, and the post-distortion oracle will favor class $1$ more, leading to a rise in type I error. Note that $C$ captures the difficulty of the classification problem: the larger $C$, the better class separation, and the easier the classification problem.

\section{Neyman-Pearson (NP) Classification Paradigm}\label{sec:np_paradigm}

One existing solution to the problem of data distortion is to collect information so as to better understand the data generation process. \textcolor{black}{For example, one might spend efforts estimating the distortion rates $\beta_0$ and $\beta_1$.} However,  such a solution is usually costly and practically infeasible. Another idea is to adjust the weight placed on each of the two types of errors in the objective function of the classical classifier. This is the cost-sensitive learning paradigm \citep{Elkan01, ZadLanAbe03}, in which users impose different costs to the two types of errors to address the issue of asymmetric error importance. However, such a method does not solve the data distortion problem, as discussed in Appendix B. To tackle the data distortion issue  and type I error control objective simultaneously, we propose to adopt the NP paradigm.

\subsection{NP Oracle Invariant to Distortion}
The NP oracle $\phi^*_{\alpha}$ arises from the famous Neyman-Pearson Lemma in statistical hypothesis testing (attached in Appendix F). Instead of minimizing $R(h)=R_0(h)\cdot\p(Y=0) + R_1(h)\cdot\p(Y=1)$ as in the classical paradigm, the NP classification paradigm aims to  mimic the NP oracle $\phi^*_{\alpha}$, where
\vspace{-0.05in}
\begin{equation}\label{eqn:np_oracle}
\phi^*_\alpha=\argmin_{\phi:~R_0(\phi)\leq\alpha}R_1(\phi)\,,
\end{equation}
in which $\alpha$ is a user-specified  upper bound on type I error. \textcolor{black}{Under the NP classification
paradigm, $\alpha$ reflects the level of a user's conservativeness towards
the type I error. In some biomedical applications, there is clear choice of
 $\alpha $, such as $.01$ and $.05$, due to either government
regulation or common practice. In social sciences applications, the choice of
$\alpha $ is more subjective. Some suggestions in choosing $%
\alpha $ can be found in  \cite{tong2016np}.}

The NP classification paradigm has three advantages: i) bypass data distortion, ii) address the class imbalance issue, and iii) control the more severe error type (typically, type I error) under a user-specified level. The third advantage is self-evident; the first two are illustrated as follows.

\begin{Theorem}\label{prop:np_invariant}
Suppose that the distortion scheme does not change the distributions for $X|(Y=0)$ and $X|(Y=1)$. The NP oracle classifier $\phi^*_{\alpha}$ defined in \eqref{eqn:np_oracle} is invariant under distortion at various rates $\beta_0$ (on class $0$) and $\beta_1$ (on class $1$), regardless of whether pre-distortion classes are balanced.
\end{Theorem}

Theorem \ref{prop:np_invariant} (proof in Appendix A) implies that in an idealized situation when one has access to the entire post-distortion population, he/she can reconstruct the NP oracle classifier as if the entire pre-distortion population is available. The rationale is that the NP oracle depends only on the conditional distributions of $X|(Y=0)$ and $X|(Y=1)$ but not on the marginal distribution of $Y$. This means that, as long as these conditional distributions do not change, the NP oracle will stay the same.

\begin{figure*}[!h]
	\centering   \includegraphics[scale=0.45]{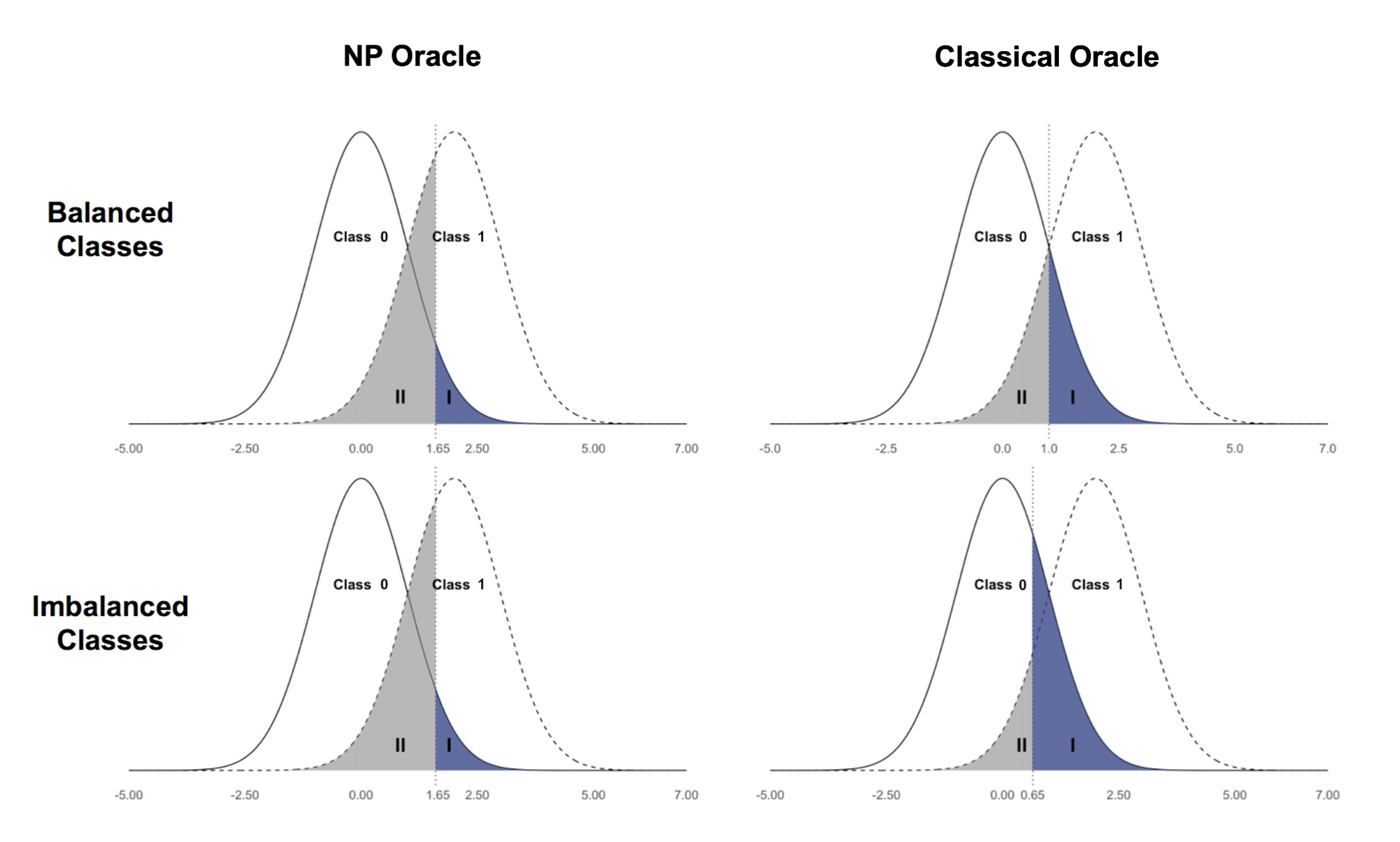} %
	\caption{\footnotesize{NP vs. Classical oracle classifiers in a Gaussian model example. The conditional distributions of $X$ under the two classes are $\mathcal N(0,1)$ and $\mathcal N(2, 1)$ respectively. Suppose that a user prefers a type I error $\le \alpha = .05$. When the two classes are balanced (i.e.,$\p(Y=0) = \p(Y=1)$), the classical oracle $\1(X > 1)$ that minimizes the risk would result in a type I error $=.159$. On the other hand, the NP oracle $\1(X > 1.65)$ that minimizes the type II error under the type I error constraint ($\leq .05$) delivers the desirable type I error. In an imbalanced situation where $2\p(Y=0) = \p(Y=1)$, while the NP oracle does not change and retains the desirable type I error, the decision boundary of the classical oracle shifts left to $.6534$ and results in a much larger type I error $=.257$.}  \label{fig:2}}
\end{figure*}

Figure \ref{fig:2} illustrates the difference between a classical oracle classifier and its NP counterpart in both balanced and imbalanced Gaussian settings. While the classical oracles are different, the NP oracle is the same in both settings.  As the type of data distortion in our study amounts to a change in the class proportion, this figure also demonstrates a contrast between a shift in decision boundary of the classical oracle and the invariance of the NP oracle under data distortion.

The main theoretical results, Theorem \ref{prop: general classical oracle under distortion} and Theorem \ref{prop:np_invariant}, do not require any parametric assumptions. We only use parametric Gaussian examples as an  illustration of these two theorems. Specifically, we use Proposition \ref{prop:1} and Figures \ref{fig:censorhip_difference} and \ref{fig:2}  to illustrate (1) the impact of data distortion on classical oracles and (2) the invariance of the NP oracles to data distortion.

Theorem \ref{prop:np_invariant} suggests that, using samples from the post-distortion population, we can train classifiers to mimic the pre-distortion NP oracle classifier due to its invariance to distortion, and thus bypass the need to estimate the data distortion scheme. \textcolor{black}{Another key implication is that one could train an NP classifier on data from a distorted population with distortion rates $\beta'_0$ and $\beta'_1$ and test the classifier on data from another distorted population with different distortion rates $\beta''_0$ and $\beta''_1$. In other words, if the training and test data undergo different censorship schemes, the NP paradigm can still be applied.}
Regarding the practical implementation of the NP paradigm, we will introduce the NP umbrella algorithm \citep{tong2016np}, which is compatible with all the scoring-type classification methods (e.g., logistic regression, support vector machines and random forest), parametric or nonparametric.

In Appendix C, we discuss a situation in which the class conditional densities of features are also changed by distortion. We derive the necessary and sufficient condition for the invariance property of the NP oracles in such a more general situation. Essentially, the general condition requires that the post-distortion class conditional density ratio is a multiple of the pre-distortion one, and that a good tail behavior is satisfied for the density ratios. We also construct concrete examples showing that these abstract generalization conditions could materialize in common model settings. Nevertheless, we choose to present the more-specific condition in Theorem \ref{prop:np_invariant} because it is transparent and easy to interpret.

\subsection{NP Umbrella Algorithm}

\begin{figure}[!h]
   \centering
   \includegraphics[trim=1in 3.7in 1in 1.2in, clip, width=1\linewidth]{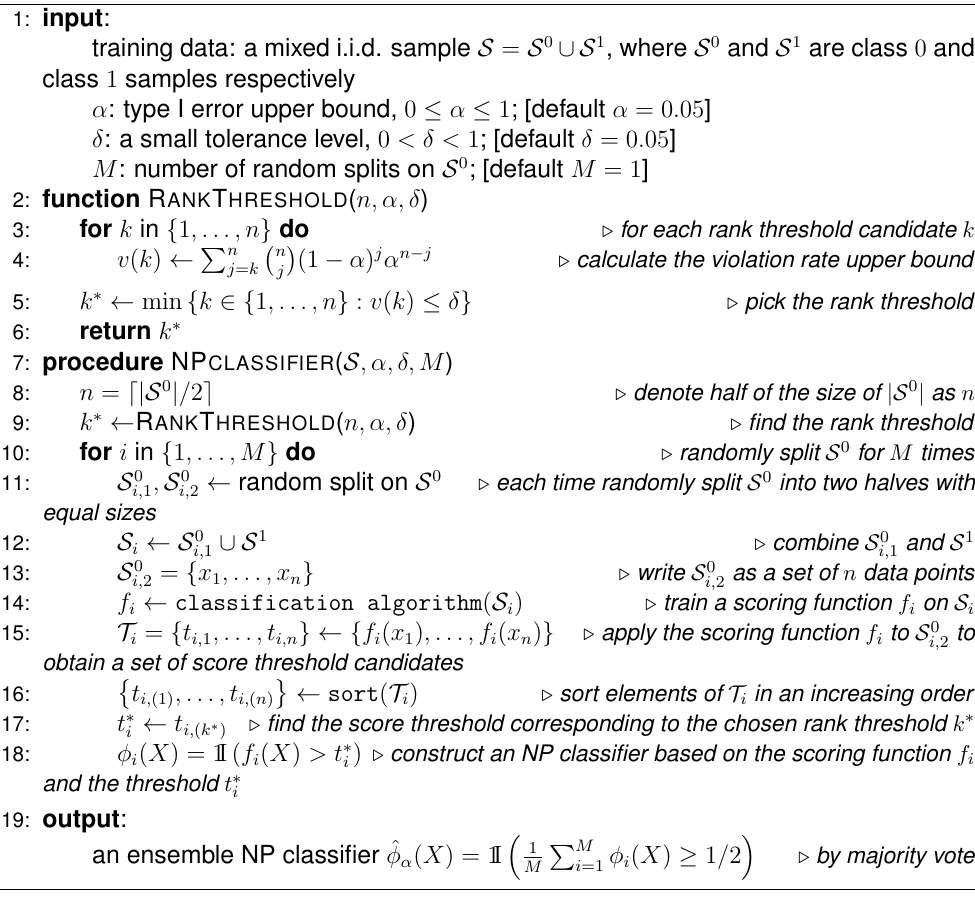}
   \caption{\footnotesize{Pseudocode for the NP umbrella algorithm adapted from \cite{tong2016np} with permission. }\label{pseudocode}}
\end{figure}

To construct a classifier under the NP paradigm, one can plug the class conditional feature densities and the threshold estimates into the NP oracle classifier suggested by the Neyman-Pearson Lemma (Appendix F). Plug-in NP classifiers have been constructed in two settings: low-dimensional \citep{Tong.2013} and high-dimensional with independent features \citep{zhao2016neyman}. However, plug-in procedures suffer from the curse of dimensionality in more general high-dimensional settings.  To make the NP paradigm more practical, \cite{tong2016np} propose an NP umbrella algorithm, a wrapper method that allows users to apply their favorite  scoring-type classification methods, such as logistic regression, support vector machines, and random forests, under the NP paradigm. Figure  \ref{pseudocode}  illustrates the pseudocode of the NP umbrella algorithm. This umbrella algorithm uses part of class $0$ data and all class $1$ data to train a scoring-function and use the left-out class $0$ data to determine a threshold for the scoring function.  To use the algorithm, a user specifies a desired upper bound $\alpha$ for the (population) type I error and an upper bound for the type I error violation rate $\delta$ (i.e., the probability that type I error exceeds $\alpha$).
Proposition \ref{prop2} and Corollary \ref{sample-based-classifier} provide a theoretical warranty for the control of type I error using the classifiers constructed based on samples.
\begin{Proposition}[adapted from \cite{tong2016np}]
\label{prop2}
	Suppose that we divide the training data into two parts, one with data from both classes $0$ and $1$ for training a base algorithm (e.g. svm, random forest and etc.) to obtain $f$ and the other as a left-out class $0$ sample for choosing the threshold. 	Applying $f$ to the left-out class $0$ sample of size $n$, we denote the resulting classification scores as $T_1, \ldots, T_{n}$, which are real-valued random variables. Then, we denote by $T_{(k)}$ the $k$-th order statistic (i.e., $T_{(1)} \leq \ldots \leq T_{(n)}$). For a new observation $X$, if we denote its classification score $f(X)$ as $T$, we can construct classifiers $\hat\phi_k (X) = \1(T > T_{(k)})$, $k\in\{1,\ldots, n\}$. Then, the population type I error of $\hat\phi_k$, denoted by $R_0(\hat\phi_k)$, is a function of $T_{(k)}$ and hence a random variable, and it holds that
	\begin{equation}
	\p\left[ R_0(\hat\phi_k) > \alpha \right] \le \sum_{j=k}^{n} {{n}\choose{j}} (1-\alpha)^j \alpha^{n-j}\,.
	\label{eq3}
	\end{equation}
	That is, the probability that the type I error of $\hat\phi_k$ exceeds $\alpha$ is under a constant that only depends on $k$, $\alpha$ and $n$. We call this probability the violation rate of $\hat\phi_k$ and denote its upper bound by $v(k) = \sum_{j=k}^{n} {{n}\choose{j}} (1-\alpha)^j \alpha^{n-j}$.
\end{Proposition}
\begin{Corollary}
\label{sample-based-classifier}
Suppose that the distortion scheme does not change the distributions for $X|(Y=0)$ and $X|(Y=1)$. The NP umbrella algorithm (with $M=1$) presented in Figure \ref{pseudocode} yields a classifier $\hat \phi$ such that $\hat\phi$ has type I error violation rate controlled, i.e.,  $\p(R_0(\hat \phi)\leq \alpha) \geq 1- \delta$, and attains the smallest type II error given a user-specified method.
\end{Corollary}
Corollary \ref{sample-based-classifier} follows from Proposition \ref{prop2}.  The proof of Corollary \ref{sample-based-classifier} can be briefly described as following. It is obvious that $v(k)$ decreases as $k$ increases.  To choose from $\hat \phi_1, \ldots, \hat \phi_{n}$ such that a classifier achieves minimal type II error with type I error violation rate less than or equal to a user's specified  $\delta$, the right order is
\begin{equation}\label{eqn:kstar}
k^* = \min\left\{k\in\{1, \ldots, n\} :   v(k) \leq \delta\right\}\,.
\end{equation}
Notice that the NP umbrella algorithm does not guarantee the type II error to be close to the oracle level, because it does not rely on assumptions of the distribution of $(X, Y)$ or the chosen classification method.

\section{Case Study}\label{sec::case study}
We present a case study regarding how to classify posts about strike events in Chinese social media. This case empirically illustrates the problem of unknown data distortion in text classification and the relevance of the NP classification approach to real-world decision making. Moreover, we demonstrate how to implement and assess various NP classification methods so that researchers of interest can adopt them.

Information regarding collective action events such as worker strikes and protests is important for citizens' participation in politics, policy implementation by governments, the accountability of political leaders, and business decisions of firms. In authoritarian countries, however, this type of information has been scarce in the public sphere because of strict government control over the mass media. The emergence of social media enables citizens to circulate information about social events and voice their opinions on political issues. This has inspired local governments, non-government organizations, firms and investors, and particularly social scientists to gather, decode and analyze the information produced on social media in authoritarian countries. However, in their endeavor to utilize information found on social media, these decision makers face the challenge of data distortion caused by extensive censorship of social media information. The NP classification approach is intended to help make use of the limited set of useful information that remains on social media to better discover and predict hidden social events.

\begin{figure}[!tb]
\centering
\includegraphics[scale=0.55]{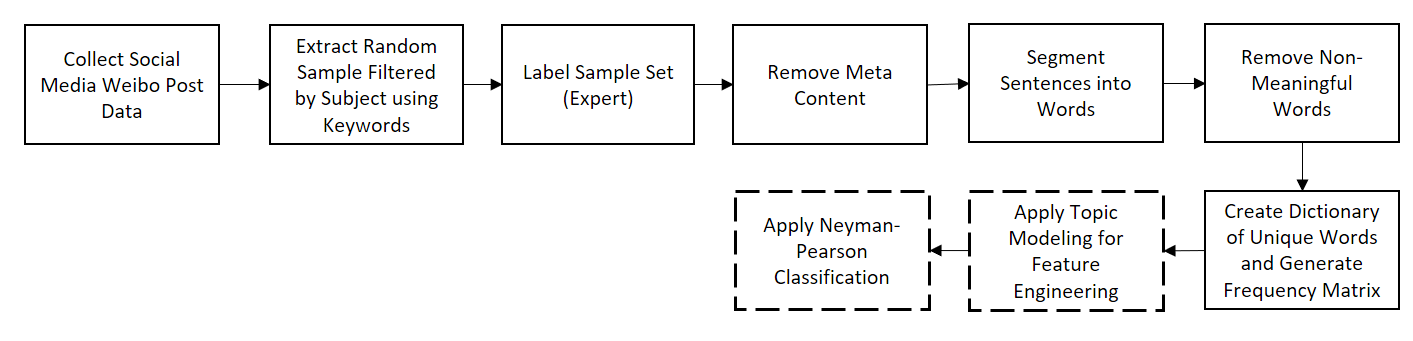}
\caption{\footnotesize{Illustration of the data processing pipeline with the pre-processing steps in the solid squares.}}
\label{fig:flowchart}
\end{figure}

In this section, we first depict the Chinese government's social media censoring strategy and explain how it fits the theoretical setup outlined in Section \ref{sec:unknown distortion}. Second, we describe our research design and data collection. Third, we detail the pipeline of data analysis including data pre-processing, feature engineering, and the implementation of each NP classification method. As a preview, Figure \ref{fig:flowchart} shows the entire chain of empirical analysis. Finally, we present the results in a baseline sample and then in four augmented samples to further illustrate the advantage of the NP classification approach.

\subsection{Data Distortion in Chinese Social Media}\label{sec:censorship}

In China, social media are typically owned by private service providers. For example, \textit{Sina Weibo}--the microblogging platform in this study-- is owned by Sina Corp., which is a company listed in NASDAQ. However, the Chinese central government controls the infrastructure based on which the social media platform operates and thus has the de facto right of censoring social media. Numerous studies have documented that the Chinese government extensively censors social media information, particularly political information that may undermine the leadership of the Chinese Communist Party, trigger large-scale collective action, and cause social unrest (\citep{chen2011internet, king2013censorship, king2014reverse}). Nevertheless, this does not mean that all politically sensitive information is censored. Using a dataset of $13.2$ billion posts published in Sina Weibo from 2009 to 2013, \cite{qin2017does}      document millions of posts published in Sina Weibo that discussed protests, demonstrations, strikes, and corruption. Based on the posting activities of users who had published this politically sensitive information, they conclude that the Chinese government allows for the circulation of some political information on social media with an intention to encourage participation and collect information for surveillance and monitoring local officials. Other studies (\citep{Lorentzen2014China, Qin2019social}) suggest that the Chinese government's strategy of censoring social media revolves around a trade-off between utilizing bottom-up information and avoiding accumulation and spread of information that may scale up existing events (e.g., protests and strikes) or spur new action. Such a tradeoff leads to the following common censorship practice: information about small local social events is not censored until a scale shift of information is detected (\citep{Bamman2012censorship, Zhu2013velocity}). In other words, when the quantity of sensitive information exceeds some threshold, censorship is triggered.

Unlike in Russia where the manipulation of online information is mostly through the deployment of bots to perform automated tasks, in China, censorship of social media is largely implemented in an ad hoc manner. The threshold of censorship depends on local social and political conditions (\citep{chen2011internet, Bamman2012censorship}).  It is well known that during the period of Congressional meetings or national celebration and in regions where social conflicts are pronounced, the Chinese government tends to tighten censorship to contain potential social unrest. This ad hoc censorship policy provides an explanation for the wide range of censorship rates estimated in existing studies. (\citep{chen2011internet, Bamman2012censorship, Fu2013assessing}).

In practice, the censorship on Chinese social media involves three additional parties other than the central government: (1) social media providers, private IT companies which implement censorship, (2) government information officers who enforce the implementation of censorship, and (3) local governments who find ways to interfere with the operation of social media. These parties may have different objectives than the central government. For example, to maintain a high level of information traffic, social media providers do not completely comply with the government's censorship demands. Moreover, the enforcement of censorship by government information officers is based on ad hoc issuing of directives, depending on the involving officers' collection and interpretation of information (\citep{chen2011internet, Zhu2013velocity}). Finally, although local governments do not have the right to censor social media, they may bribe employees of social media providers to delete information that may reflect negatively on them.

The above characteristics regarding censorship make the Chinese social media an ideal setting to study the problem of classification in the presence of data distortion and the NP classification methods as a solution to the problem. A decision maker who wishes to extract useful information about certain issues or events from post-censorship social media posts faces the problem of data distortion as we formulate in the previous sections. The quantity-based censorship suggests that the features of information in the relevant class are likely to remain stable despite that censorship significantly reduces the quantity of this type of information. Therefore, the key assumption under which the invariance property of the NP oracle classifier is approximately true. Importantly, the ad hoc nature of censorship and the involvement of multiple parties in its implementation render the actual censorship scheme highly volatile and unpredictable. It is practically infeasible for a decision maker to infer the rate of data distortion due to censorship.

\subsection{Data Collection and Research Design}

For this study, we collected public user posts related to sensitive social issues from the microblogging site \textit{Sina Weibo}. Through a third-party content crawling agency, we obtained a dataset of approximately $10$ million raw posts about public issues and social events in $2012$. We are interested in classifying posts about the subject ``worker strikes."  We focus on strikes for several reasons. First, the number of strikes in China has surged in the last decade, and strikes have become an important form of worker movement (\citep{ChinaLaborBulletin2011, ChinaLaborBulletin2018}). Accurately identifying strike events in a timely manner is important for a wide range of decision makers, including governments, firms, and social scientists. Second, as an indicator of collective action, posts about strikes are prone to censorship. Third, the degree of censorship of posts about strikes varies across regions and over time. For instance, censorship tends to be more intense towards the end of the year when workers' yearly compensation is due and in regions where economic conditions are worse and unemployment rates have increased. As explained below, this variation provides a partial test of the assumptions that entail the application of our proposed NP classification methods as well as an opportunity to demonstrate the advantage of the NP classification approach.

\begin{CJK*}{UTF8}{gbsn}

We extract a subset of posts filtered according to a pre-selected list of keywords.\footnote{The filter for strike includes the following list of keywords, which commonly appear with the subject:	``罢工(worker strike)", ``工潮(worker strike)",``罢市(shopkeeper strike)", ``罢课(class boycott)", ``罢驶(stop driving)", ``罢驾(stop driving)", ``罢运(transportation worker strike)".} This filtering generates $221,229$ posts linguistically relating to strikes. From this dataset, we extract a random sample of $2,500$ posts that were published in the first quarter of 2012 and were originated from Guangdong -- a coastal province where strike incidence occurred most frequently among all provinces in China during the sample period. This sample serves as a baseline for our data analysis as well as an illustration of various NP classification methods. We then extract three random samples of the same size ($2,500$ posts) in the 2nd, 3rd, and 4th quarters of 2012 from Guangdong, respectively. We will apply an NP classifier trained in the baseline sample to these three samples in other periods. Good performance (in terms of controlling type I error) of this classifier across different samples provides suggestive evidence on the stable distribution of features in the presence of data distortion. Finally, we select a random sample of $2,500$ posts from all the posts originated from three inland provinces--Gansu, Qinghai, and Xinjiang--during the entire year of 2012. Evidence shows that these three provinces were among regions where censorship on social media was most intense (\citep{Bamman2012censorship}). Therefore, information that can be used to discover and predict strikes is expected to be scarce in these provinces. Again, we will apply the NP classifier trained in the baseline sample to this non-Guangdong sample. If this classifier performs well on the new sample, decision makers can use a classifier trained in an environment with relatively abundant information to overcome the challenge of classification in an information-scarce environment where labelling of posts is likely to be much more costly and cannot be done in a timely manner. This is a potential advantage of the NP classification approach in its ability to transfer knowledge from one domain to another.
\end{CJK*}

\subsection{Data Pre-processing}

We now describe how we process the unstructured raw Sina Weibo posts so that they can be fed to learning algorithms. The first step is to generate post labels. A decision maker's interest is to learn strike events which are a form of workers' collective action and reflect ongoing social and economic problems. Labeling posts according to the decision makers' interest turns out to be non-trivial for two reasons. First, in terms of substance, some posts related to strikes are about events in history or in other countries without implications for current events. Second, linguistically, the word ``strike" is widely used in many different contexts, literally and metaphorically. For example, in Chinese, in the sentence ``my computer / my cell phone is on strike," ``strike" means "has stopped working." In the sentence ``A person's body / brain is on strike," ``strike" means ``is not functioning normally." This type of linguistic ambiguity exists in many languages. We specified a set of rules to capture these subtleties.

As a trial, we outsourced the labeling task to Amazon Mechanical Turk. Despite the active responses, the label quality was subpar, having many errors and inconsistencies. Realizing the difficulty of the task, we switched to expert labeling. We hired two Chinese-speaking experts to manually categorize the raw posts into ``strike related" (class $0$) and ``strike unrelated" (class $1$). Class $0$ are posts about worker strikes, including student strikes, taxi driver strikes, and merchant strikes, whereas class $1$ posts contain the keyword ``strike" but are using the word metaphorically to describe the malfunctioning of computers, elevators or other objects.

After trial and error, the two experts achieved high quality and consistent labeling in several trial samples. They then labeled the five aforementioned random samples: the baseline sample (GD-Q1), the three other samples in Guangdong after the first quarter in 2012 (GD-Q2, GD-Q3, and GD-Q4), and the sample outside of Guangdong (NGD). Overall, among the $12,500$ posts in these five samples, $3,237$ posts are labeled as ``strike related" (Class 0) and $9,263$ as ``strike unrelated" (Class 1).

To decipher which Chinese characters form meaningful words, we apply \textit{The Stanford Segmenter} \citep{tseng2005conditional}, which uses a Chinese treebank (CTB) segmentation model and breaks down input messages into disjointed words. After removing non-meaningful stop words, we create a dictionary of unique words and generate a frequency matrix that counts the number of times each word appears in each post, based on the dictionary. The \textit{strikes} matrix, containing $12,500$ rows (posts) and $34,968$ columns (features), is used to engineer features in topic modeling.

\subsection{Feature Engineering}

In the pre-processed  \textit{strikes} dataset, the size of vocabulary dictionaries is much larger than the number of posts. This high-dimensional problem can be handled with various techniques. For example, one can use marginal screening methods such as sure independence screening \citep{Fan.Lv.2008}, nonparametric independence screening  \citep{Fan.Feng.Song.2011} and the Kolmogorov-Smirnov (KS) test, interaction screening methods \citep{Hao.Zhang.2014,Fan.Kong.Li.Zheng.2015}, the forward stepwise selection, shrinkage methods such as  LASSO \citep{Tibshirani.1996} and SCAD \citep{Fan.Li.2001},   or dimension reduction methods such as principal component analysis.

These methods, however, all overlook the semantic structures possessed by corpora datasets. %
Thus, we adopt Latent Dirichlet Allocation (LDA) \citep{blei2003latent, teh2007collapsed, grimmer2013text}, which is a popular generative probabilistic model designed for large corpora. In this model, documents (posts) are represented as random mixtures over latent topics and each topic is represented as a distribution over words. %
We train the LDA model using the \texttt{R} package \texttt{topicmodels} and select ``Gibbs sampling" as the fitting method.  With a pre-determined $K$, we extract $K$ topics that serve as new features. The posterior distribution over these $K$ topics in each document will be the feature values.

\subsection{Results}

In this subsection, we present the main results of the analysis which follows the pipeline depicted in Figure \ref{fig:flowchart}. Alongside the results, we discuss their real-world implications. We also address several nuanced technical issues that are important for the implementation of classification methods, hoping to provide quantitative social scientists some implementation guidelines to analyze their classification problems in various empirical settings.

\subsubsection{Topic Modelling}\hfill

 In the use of LDA for feature engineering, specifying the number of topics $K$ is essential. We use a \textit{stability} criterion to select $K$. Concretely, for a candidate $K$, we randomly select half of the posts to apply LDA. This process is repeated $50$ times. Every time, LDA outputs $K$ topics. Each document is represented by posterior probabilities over these $K$ topics, and each topic is represented by posterior probabilities over the vocabulary dictionary. We look at the top $20$ keywords that have the largest posterior probabilities in each of the $K$ topics.  Based on these words, we decide whether a topic is truly related to the subject. We consider the number of topics $K$ to be suitable if over $50$ repetitions, the proportions of relevant topics have low variance. For illustrative purpose, we compare $K=5$ and $K=10$.

Table \ref{table:strikekeywords} lists the top $20$ keywords for each topic in one repetition when $K=10$, using the entire dataset of 12,500 strike posts. Even a casual reader (of the Chinese characters or their corresponding English translation) will recognize the 4th, 6th and 10th topics as about actual worker strikes. In particular, the 4th topic is mostly about students boycotting classes, evident from the keywords ``school," ``student," ``teacher," ``student strike," and ``demonstration." The 6th topic is about worker strikes in firms, evident from the keywords ``company," ``employee,"  ``wage," ``protest," ``collective," ``factory," and ``staff." The 10th topic is about strikes in the transportation sector, evident from the keywords ``strike," ``driver,"  ``vehicle," ``taxi," ``public transportation," ``collective," ``road," ``bus," and ``traffic." The remaining $7$ topics are irrelevant. Thus, in this repetition, the proportion of relevant topics is $3/10$.  Over the $50$ repetitions, we calculated the variances of these proportions.  In this regard, $K=5$ and $10$ output variances .0037 and .0018 respectively. By the stability criterion, we prefer $K=10$ .

\begin{CJK*}{UTF8}{gbsn}
One interesting observation is that, for a different choice of $K$, the feature words in a specific topic may contain different information. For example, the topic regarding ``strikes in the transportation sector (topic 10 in Table 1)" appears both when $K=5$ and when $K=10$. In addition to relating to the subject, the topic keywords also contain information about the location of the events, which is valuable for decision making. However, when $K=5$, only one location ``汕头" (Shantou) appears as a feature in the selected topic; whereas when $K=10$, two locations, ``汕头" (Shantou) and ``江门" (Jiangmen), appear. To investigate the cause of this difference, we manually read through the $2,500$ posts we selected from GD-Q1. Of them, $230$ posts are about strikes in ``Shantou" and $161$ posts are about strikes in ``Jiangmen." We suspect that it is the relatively low frequency of ``Jiangmen" that makes it vanish as a feature in the selected topic when $K=5$. Thus, choosing a larger $K$ may have the advantage of capturing a greater amount of valuable information.  In the remaining part of the paper, we set $K=10$ unless otherwise specified.
\end{CJK*}

It should be noted that, in this study, we choose $K=5$ or $K=10$ simply for an illustrative purpose. Practitioners can select a set of desirable $K$'s based on their domain knowledge, time constraint, and financial budget.

\begin{CJK*}{UTF8}{gbsn}
	\begin{table*}
		\fontsize{6pt}{9pt}\selectfont
		\begin{center}
			\setlength\tabcolsep{2pt}
			\begin{tabular}{ccccccccccc}
				\hline
				topic 1&去 &今天 &吃 &明天&做&上班&爱&睡觉&今晚&偷笑\cr
				&go&today&eat&tomorrow&do&work&love&sleep&tonight&smirk\cr
				&下午&然后&鼻屎& 挖&回来&回家&睡&买&晚上&累\cr
				&afternoon&afterwards&mucus&pick&come back&go home&sleep&buy&night&tired\cr
				\hline
				topic 2&罢工&电脑&手机&打&发现&居然&电话&突然&开&发\cr
				&strike&computer&cellphone&beat&realize&surprisingly&telephone&sudden&open&send\cr
				&换&最近&直接&结果&系统&部&问题&彻底&博&家里\cr
				&change&recently&directly&consequence&system&part&problem&thoroughly&broad&home\cr
				\hline
				topic 3&罢工&天&说&现在&小&点&今天&后&下&三\cr
				&strike&day&speak&now&small&bit&today&after&down&three\cr
				& 前&真的&早上&知道&去&分钟&走&思考&一直&真是\cr
				&before&really&morning&know&go&minute&walk&think&always&indeed\cr
				\hline
				topic 4& 年&工人&罢课&中国&上&月&工会 &游行&政府&老师\cr
				&year&worker&student strike&China&previous&month&union&demonstration&government&teacher\cr
				&学生 &学校&美国&国家&生活&组织&人民&中&领导&举行\cr
				&student&school&United States&country&life&organization&people&in&leader&hold\cr
				\hline
				topic 5& 能 & 让 &可以&时候&没有&还是&这个&种&上&觉得\cr
				&can&let&may&time&none&still&this&plant&up&feel\cr
				&希望&出来&里&工作&身体&感觉 &太阳&还有&出&一定\cr
				&hope&come out&inside&work&body&feel&Sun&still&out&definitely\cr
				\hline
				topic 6& 公司&员工&事件&工资&工作&抗议&对&罢市&中&发生\cr
				&company&employee&event&wage&work&protest&right&shopkeeper strike&in&happen\cr
				&新闻& 集体&问题&三&分享&月日&工厂&人员& 新&政府\cr
				&news&collective&problem&three&share&month-date&factory&staff&new&government\cr
				\hline
				topic 7& 罢工&抓 &狂&泪 &系&今日&可怜&地&生病&衰\cr
				&strike&clutch&crazy&tear&be&today&pity&ground&sick&unfortunate\cr
				&抓狂&泪泪& 天气&委屈&话 &鄙视&住 &空调&搞到&甘\cr
				&go crazy&tears&weather&be wronged&word&despise&reside&air-conditioner&get&this\cr
				\hline
				topic 8& 罢工&想&人&玩 &哈哈&事& 找&为什么&心情&回复\cr
				&strike&think&human&play&haha&thing&find&why&mood&reply\cr
				&汗&看到&伤心&事情&很多&闹钟&放假&二&个人&双\cr
				&sweat&see&sad&thing&many&alarm clock&holiday&two&individual&pair\cr
				\hline
				topic 9& 罢工&次 &开始 &时间&过&已经&终于&最后&继续&嘻嘻\cr
				&strike&time&begin&time&pass&already&finally&at last&continue&LOL\cr
				&第一&对&周 &所有&下&第二 &机场&地方&草草&完全\cr
				&first&right&week&all&down&second&airport&place&hasty&completely\cr
				\hline
				topic 10& 罢工&司机 &车 &出租车&的士&公交&集体&小时 &汕头&路\cr
				&strike&driver&car&taxi&taxi&public transportation&collective&hours&Shantou&road\cr
				&钱&没有&回&公交车&江门& 半&交通&事&全部&广州\cr
				&money&none&back&bus&Jiangmen&half&traffic&thing&all&Guangzhou\cr
				\hline
			\end{tabular}
			\caption{\footnotesize{top $20$ keywords for ten topics from one repetition on the entire \textit{strikes} dataset. The English translation of some keywords in Topic 7 are based their Cantonese meaning.}}
			\label{table:strikekeywords}
		\end{center}
	\end{table*}
\end{CJK*}
\subsubsection{NP Classification in the Baseline Sample}\label{section:settings}\hfill

Fixing $K=10$ in LDA, we apply both the classical and NP classification algorithms to the baseline dataset GD-Q1. The NP algorithms are implemented through the \verb+R+ package \texttt{nproc} (also available in \verb+Python+).  To better demonstrate the performance of NP classifiers, we implement three settings.%

\begin{itemize}
\item{\sf{Setting 1}:} We randomly split GD-Q1 into training and test sets of equal sizes (half of class $0$ and half of class $1$ data in training) $100$ times. Hence, the class $0$ proportion in a training set is the same as that in a test set. We set NP parameters: $\alpha = .2$ and $\delta= .3$.

\item{\sf{Setting 2}:} We randomly split class $0$ data into three folds of equal sizes, and split class $1$ data into two halves. We take $1/3$ (one fold) class $0$ data and $1/2$ class $1$ data as the training set and use the other $2/3$ class $0$ data and the other half class $1$ data as the test set. Thus, the class $0$ proportion in the training set is half as much as in the test set. We again repeat the experiment $100$ times. We set NP parameters: $\alpha = .2$ and $\delta= .3$.

\item{\sf{Setting 3}:} The same as in {\sf{Setting 1}}, except that we now set NP parameters: $\alpha = .1$ and $\delta= .3$.
\end{itemize}

 In each training set, we run LDA ($K=10$) and construct a transformed training set which utilizes the learned topics as new features and the posterior probabilities over these topics as feature values. We then train classifiers based on the transformed training datasets. Type I and type II errors are calculated using the corresponding transformed test set.  The classification methods implemented include the classical versions of penalized logistic regression (PLR), naive bayes (NB), support vector machines (SVM),  random forest (RF) and sparse linear discriminant analysis (sLDA), together with their NP counterparts with corresponding parameters (e.g., $\alpha = .2$ and $\delta= .3$ for {\sf Setting 1} and {\sf Setting 2}; $\alpha = .1$ and $\delta= .3$ for {\sf Setting 3}).

Table \ref{table:2}  summarizes the average type I and type II errors  in {\sf{Setting 1}} using the above classification methods under the classical approach (odd columns) and the NP approach (even columns, named with a prefix NP) over all $100$ repetitions. Notably, all the classical methods produce a large type I error and a small type II error, with Naive Bayes being the most extreme one, where the type I error is $1$. This is in part caused by the relatively large size of Class 1 in the training dataset. By contrast, all NP methods successfully control the type I error within the target level, while producing a larger type II error than that of the classical methods. This means that a decision maker using the NP methods can more accurately discover true information about strike events at the cost of screening some extra irrelevant information. In the current study, missing a strike-related post (class $0$) may lead to delayed government responses, oversight in business decisions, and under-estimates of the strike incidence frequency in social studies. It is particularly costly when a hidden event may compound into a large scale issue and spread to other regions. Generally, a decision maker cares more about type I error than type II error. The larger type II error associated with the NP classifiers implies that an excessive amount of irrelevant information has been collected and another round of screening may be needed. The cost of such further screening appears insignificant \citep{qin2017does}. Overall, the NP classifier is preferable in many real-world applications.

\begin{table}[tp]
\small{
\begin{center}
\begin{tabular}{ c c c c c c c c c c c}
\hline
 Error rates & PLR&NP-PLR & NB &NP-NB & SVM & NP-SVM & RF & NP-RF&sLDA&NP-sLDA
\\
\hline
 type I & .914 & .196 &1 &.193 & .816 &.179&.684 &.184&.825&.194\\
 type II & .005 & .427 &0 &.482 &.014&.598&.047&.502&.014&.423\\
 \hline
 \end{tabular}
\caption{\footnotesize{Average error rates with $\alpha=.2$, $\delta=.3$ for the strike dataset over $100$ repetitions, under {\sf{Setting 1}}.}}
\label{table:2}
\end{center}}
\end{table}

\begin{table}[!t]
\small{
\begin{center}
\begin{tabular}{ c c c c c c c c c c c}
\hline
 Error rates & PLR&NP-PLR & NB &NP-NB & SVM & NP-SVM & RF & NP-RF&sLDA&NP-sLDA
\\
\hline
 type I & .965 & .184 &1 &.183 & .918 &.166&.822 &.169&.872&.185\\
 type II & .002 & .498 &0 &.571 &.005&.732&.023&.588&.010&.494\\
 \hline
 \end{tabular}
\caption{\footnotesize{Average error rates with $\alpha=.2$, $\delta=.3$ for the strike dataset over $100$ repetitions, under {\sf{Setting 2}}.}}
\label{table:3}
\end{center}}
\end{table}

\begin{table}[tp]
\small{
\begin{center}
\begin{tabular}{ c c c c c c c c c c c}
\hline
 Error rates & PLR&NP-PLR & NB &NP-NB & SVM & NP-SVM & RF & NP-RF&sLDA&NP-sLDA
\\
\hline
 type I & .912 & .095 &1 &.089 & .824 &.084&.690 &.083&.826&.097\\
 type II & .006 & .659 &0 &.733 &.014&.806&.047&.740&.014&.651\\
 \hline
 \end{tabular}
\caption{\footnotesize{Average error rates with $\alpha=.1$, $\delta=.3$ for the strike dataset over $100$ repetitions, under {\sf{Setting 3}}.}}
\label{table:4}
\end{center}}
\end{table}

Table \ref{table:3}  summarizes the average type I and type II errors in {\sf{Setting 2}}, in which the class $0$ proportion in the training set is half as much as its proportion in the test set. This mimics the real life scenario when censorship of the more-sensitive information is tightened, resulting in more scarce relevant information (a smaller class $0$) in the observed data. According to our previous theoretical discussion, such more stringent censorship would shift the decision boundary of the classical oracle classifier more drastically, worsening type I error of the classical classification methods. For example, PLR produces a type I error of $.965$, which is larger than $.914$ -- its counterpart in {\sf{Setting 1}}. By contrast, the NP oracle is unaffected by data distortion, and  NP-PLR has a type I error controlled below the pre-specified $\alpha=.2$ in both {\sf{Setting 1}} and {\sf{Setting 2}}. This phenomenon is consistent across all the five methods we implemented.

Table \ref{table:4}  summarizes the average type I and type II errors of these methods in {\sf{Setting 3}}, which is the same as in {\sf{Setting 1}} except that we now use a new set of parameters $\alpha=.1$, $\delta=.3$. This second set of parameters is chosen to represent a scenario when decision makers face a higher cost of missing a strike event and wish to impose more stringent control over type I error. Tables \ref{table:2} and \ref{table:4} demonstrate that, across different NP classifiers, type I errors are uniformly controlled under the target level. In particular, when the upper bound of type I error is reduced from $(\alpha=.2)$ to $(\alpha=.1)$, type I errors of the NP classifiers are reduced below the new target level $.1$. These observations suggest that the NP methods provide an instrument for decision makers to fine-tune the target level of type I errors according to circumstances.

In summary, the parameters $\alpha$ and $\delta$ in NP classification methods govern the trade-off between type I and type II errors, and the balance of this trade-off depends on the decision maker's objective and resources available. In the \textit{strikes} example, the consequence of making type I errors is severe -- it could threaten government stability, jeopardize a politician's career, or mislead business decisions, whereas the cost of dealing with type II errors is small. Considering this preference for controlling type I error, together with the data distortion problem, it is highly valuable to use classification methods under the NP paradigm rather than under the classical paradigm. In Appendix D, we also demonstrate how sparsity-inducing methods, such as NP-sLDA, help select meaningful topics, so that our approach achieves both good prediction performance and good interpretability.

\subsubsection{Knowledge Transfer: NP Classifiers across Datasets}\hfill

In practice, decision makers often need to make decisions quickly. This time constraint sometimes restricts the amount of information available for developing predictive algorithms. For instance, a decision maker wants to assess the work conditions of a region (e.g., province) in April using social media posts. However, the number of relevant posts may be too small to train an effective classifier, or there might not be enough time and resources to hire experts to label posts. If this decision maker could use a classifier trained with data collected in the first quarter of the year, his or her learning would be more efficient and timely. Similarly, in a region where information related to worker strikes is scarce because of extensive censorship or limited supply, data analysis based on machine learning will benefit substantially from information collected in other regions with less censorship or more information supply.

\begin{CJK*}{UTF8}{gbsn}
The above discussion conveys the notion of \textit{knowledge transfer}, which is implied by the invariance property of the NP classification paradigm. We now examine its validity empirically. We use classifiers trained on all posts in the baseline sample (GD-Q1) to classify posts in other datasets (GD-Q2, GD-Q3, GD-G4, and NGD). From the previous section (recall Tables \ref{table:2}, \ref{table:3} and \ref{table:4}), we find that NP-sLDA performs the best among all methods we compared in terms of type II errors. Thus, in this section, we focus on NP-sLDA only. In Table \ref{table:5}, we first present the results with parameters $\alpha=.1$ and $ \delta=.3$. Note that the type I errors are slightly larger than the target control level $\alpha=.1$. Nevertheless, this does not mean the failure of applying the classifiers trained in GD-Q1 because some regional and time-varying features are specific to a dataset and cannot be used for learning in other datasets. For example, `` 汕头" (Shantou) is a prefecture in Guangdong province where taxi-driver strikes occurred multiple times in the first quarter of 2012, and thus this locality appeared as a pronounced feature in topics selected from the baseline dataset. Unless the strike events in this locality lasted for a long period and became national, we would not expect it to appear as an important feature for data from samples in other periods or from non-Guangdong provinces. In other words, we expect that the underlying populations over time or in different regions are not identical.
\end{CJK*}

Being aware of the above learning barrier caused by features that are specific to a particular sample, we propose to have a smaller tolerance level to control the desirable type I error. In particular, we trained the classifier using $(\alpha=.1, \delta=.05)$. Table \ref{table:6} presents the results under this new criterion. In contrast to the results in Table \ref{table:5}, the type I error is now well controlled under the target level $.1$.

The above results demonstrate that, armed with NP classifiers, a decision maker, who is constrained by available information and time, can leverage information collected from previous periods or in circumstances where useful information was not severely censored. Of course, this knowledge transfer is feasible only if the post-censorship feature distributions remain sufficiently stable across datasets. Therefore, the results in Tables  \ref{table:5}  and  \ref{table:6}  provide suggestive evidence that censorship does not distort feature distributions that are important for the algorithm's learning process. In other words, the assumption that warrants the invariance property of the NP oracle classifier is partially justifiable in the current empirical setting, although this assumption is not directly testable because uncensored data are not available. As mentioned in the introduction, our practice of using the NP algorithm to handle the data distortion problems differs from any existing practice in domain adaptation in that we do  not use any data (labeled or unlabeled) on the target domain in the algorithm training process.

\begin{table}[tp]
\small{
\begin{center}
\begin{tabular}{ c c c c c}
\hline
 Error rates & Guangdong-Q2&Guangdong-Q3 & Guangdong-Q4 &non-Guangdong \\
\hline
 type I & .133 & .141 &.109 &.106 \\
 type II & .558 & .563 &.516 &.533\\
 \hline
 \end{tabular}
\caption{\footnotesize{Average error rates with $\alpha=.1$, $\delta=.3$ for posts from GD-Q2, GD-Q3, GD-Q4 and NGD.}}
\label{table:5}
\end{center}}
\end{table}

\begin{table}[tp]
\small{
\begin{center}
\begin{tabular}{ c c c c c}
\hline
 Error rates & Guangdong-Q2&Guangdong-Q3 & Guangdong-Q4 &non-Guangdong \\
\hline
 type I & .094 & .100 &.087 &.078 \\
 type II & .642 & .654 &.622 &.611\\
 \hline
 \end{tabular}
\caption{\footnotesize{Average error rates with $\alpha=.1$, $\delta=.05$ for posts from GD-Q2, GD-Q3, GD-Q4 and NGD.}}
\label{table:6}
\end{center}}
\end{table}

\begin{table}[tp]
\scriptsize{
\begin{center}
\begin{tabular}{ c |c c |c c |c c| c c |c c}
\hline
&\multicolumn{2}{|c|}{NP-PLR}&\multicolumn{2}{|c|}{NP-NB}&\multicolumn{2}{|c|}{NP-SVM}&\multicolumn{2}{|c|}{NP-RF}&\multicolumn{2}{c}{NP-sLDA}\\
\hline
 Error rates  & GD-Q1&GD-ALL & GD-Q1 &GD-ALL &GD-Q1 &GD-ALL& GD-Q1 & GD-ALL&GD-Q1&GD-ALL\\
\hline
 type I & .095 & .094 &.089 &.093 & .084&.090&.083&.091&.097&.093\\
 type II &.659 &  .420&.733 & .491&.806&.587&.740&.476&.651&.425\\
 \hline
 \end{tabular}
\caption{\footnotesize{Average error rates using NP-methods with $\alpha=.1$, $\delta=.3$ over $100$ repetitions. A comparison between using GD-Q1 and all data from Guangdong.}}
\label{table:7}
\end{center}}
\end{table}

\begin{table}[tp]
\small{
\begin{center}
\begin{tabular}{ c c c}
\hline
 Error rates & trained over GD-Q1 only&trained over all data from GD \\
\hline
 type I & .078& .059  \\
 type II & .611 & .407 \\
 \hline
 \end{tabular}
\caption{\footnotesize{Average error rates with $\alpha=.1$, $\delta=.05$ for posts from NGD, using classifier NP-sLDA trained on GD-Q1 only and trained on all data from GD (including GD-Q1, GD-Q2, GD-Q3 and GD-Q4), respectively.}}
\label{table:8}
\end{center}}
\end{table}
\qquad \\
\subsubsection{Knowledge Accumulation: NP Classifiers with Enlarged Training Data}\hfill

In reality, a decision maker often accumulates information from the past. In view of the invariance property of the NP methods, this accumulated information can be used to facilitate learning if the feature distributions remain stable. We illustrate this point in the current case study. We first repeat {\sf Setting 3} using all posts from the Guangdong province, and compare the results with those in Table \ref{table:4}. In the comparison (presented in Table \ref{table:7}), GD-Q1 recollects the results in Table \ref{table:4}, and GD-ALL reports the results obtained from information in Guangdong over the entire four quarters. Clearly, when we use the larger dataset, the type II error decreases, while the type I error remains under control at $.1$. Furthermore, we include all posts from Guangdong as the training data, and test on the NGD dataset. We keep the parameters $(\alpha=.1, \delta=.05)$ the same for comparison with Table \ref{table:6}. Table \ref{table:8} shows that,  with type I error under control using NP-sLDA, the larger size of the training data decreases the type II error one would achieve on the posts from non-Guangdong, even if the underlying population distributions in the GD and NGD datasets can be different.

\section{Conclusion}\label{sec:conclusion}

Digital texts have become an important source of data for social scientists. With increasing sophistication in text mining to discover social events and to predict social behaviors, accurate classification of textual data for specific purposes is key to successful empirical analysis. However, while a wide range of textual analysis and machine learning techniques have been introduced into the social sciences \citep{grimmer2013text,wilkerson2017large,gentzkow2017text}, the problem of data distortion has received relatively little attention. Being a fundamental data generation issue in statistical analysis, data distortion can cause serious problems in sampling, inference, and prediction. The current paper is among the first efforts to study data distortion problems in the context of classifying large-scale textual data. Theoretically, we show that in the presence of unknown data distortion, the classical oracle classifier cannot be recovered even when the entire post-distortion population is available. By contrast, the NP oracle classifier is unaffected by data distortion. Practically, we study a case in which a decision maker classifies posts about worker strikes obtained from Sina Weibo -- a leading Chinese microblogging platform that is subject to government censorship. We demonstrate that when one type of classification error (e.g., type I error) is dominantly important, the NP classification algorithms allow users to control that type of error below a pre-specified level. Although our problem setup involves the distortion parameters, our objective is not to estimate them, but to bypass the estimation needs for prediction purpose. In other words, we target a prediction problem rather than an inference problem.  Our approach is to construct classifiers under the NP paradigm, and the theoretical underpinning behind this construction is the invariance property of the NP oracle classifier. It is important to note that the NP classification approach we propose is not specific to text classification. Instead, it can be used to handle more-general classification problems in the big data era when classification errors are asymmetric in importance. Plausible applications include control of epidemic diseases, crime detection, social surveillance, and monitoring risky financial decisions, among many others.

\section*{Acknowledgement}
The authors would like to thank the editor, associate editor, two statistical content referees and the referee for reproducibility, for many constructive comments which have greatly improved the paper. We would also like to thank Professor Jingyi Jessica Li for rounds of thoughtful discussions and suggestions, and the seminar participants at UCLA.  This work was partially supported by National Science Foundation grant NSF DMS 1613338.
\clearpage
\bibliographystyle{unsrtnat}
\bibliography{NP_data_distortion}

\clearpage

\appendix

\section{Proofs}\label{proof:thm12}

\subsection{Proof of Theorem 1}
\begin{proof}
Recall that the (classical) oracle classifier regarding the pre-distortion population is $h^*(x) = \1(\eta(x) > 1/2)$, where the regression function $\eta(x) = \E(Y|X=x)$ can be calculated as
$$
\eta(x) = \frac{\pi_1 f_1(x)/f_0(x)}{\pi_1 f_1(x)/f_0(x) + \pi_0}\,.
$$
Therefore, $h^*(x) = \1\left( \frac{f_1(x)}{f_0(x)}> \frac{\pi_0}{\pi_1}\right)$.
When distortion with rates $\beta_0$ and $\beta_1$ is applied to class $0$ and class $1$ respectively, the class proportions become $\pi_0^{(\beta_0, \beta_1)}$ and $\pi_1^{(\beta_0, \beta_1)}$ which are defined as
\begin{align*}
&\pi^{(\beta_0, \beta_1)}_{0} = \frac{(1-\beta_0) \pi_0}{(1-\beta_0) \pi_0 + (1-\beta_1)\pi_1}\,, \text{  }\cr
&\pi^{(\beta_0, \beta_1)}_{1} = \frac{(1-\beta_1) \pi_1}{(1-\beta_0) \pi_0 + (1-\beta_1)\pi_1}\,,
\end{align*}
while class conditional densities remain $f_0$ and $f_1$. Then, the  oracle classifier regarding the post-distortion population is to replace $\pi_0$ and $\pi_1$ in $h^*$ by $\pi^{(\beta_0, \beta_1)}_{0}$ and $\pi^{(\beta_0, \beta_1)}_{1}$ respectively:
$$h^*_{(\beta_0, \beta_1)}(x) = \1\left( \frac{f_1(x)}{f_0(x)}> \frac{\pi_0^{(\beta_0, \beta_1)}}{\pi_1^{(\beta_0, \beta_1)}}\right) = \1\left(\frac{f_1(x)}{f_0(x)}> \frac{1-\beta_0}{1-\beta_1}\cdot\frac{\pi_0}{\pi_1}\right)\,.$$
\end{proof}

\subsection{Proof of Theorem 2}

\begin{proof}
The constrained optimization program (4) in the main text that defines $\phi^*_{\alpha}$ does not involve  the class priors $\pi_0 = \p(Y=0)$ and $\pi_1 = \p(Y=1)$, so $\phi^*_{\alpha}$ does not depend on $\pi_0$ or $\pi_1$. Now suppose distortion with rates $\beta_0$ and $\beta_1$ is imposed on class $0$ and class $1$ respectively, then the post-distortion population have class $0$ proportion $[(1-\beta_0)\pi_0]/[(1-\beta_0)\pi_0 + (1 -\beta_1) \pi_1]$ and class $1$ proportion $[(1-\beta_1)\pi_1]/[(1-\beta_0)\pi_0 + (1 -\beta_1) \pi_1]$, while keeping the distributions of $X|(Y=0)$ and $X|(Y=1)$ unchanged. Since distortion at rates $\beta_0$ and $\beta_1$ only changes class proportion, which NP oracle does not depend upon, the NP oracle is invariant under distortion.
\end{proof}

\section{Cost-sensitive (CS) Learning}\label{sec:cost-sensitive learning}

An insight from studying the classical classification paradigm is  that the relative size of classification errors comes largely from the relative weights placed on type I and type II errors in the objective function. So a natural candidate to adjust classification errors is to change the weights. This is the so-called cost-sensitive (CS) learning paradigm, in which users impose costs $C_0$ and $C_1$ to type I and type II errors, respectively. On the population level, instead of minimizing the overall classification error $R(\cdot)$, one minimizes the CS learning objective:
\begin{eqnarray}\label{eqn:cost-sensitive}
\min_h R^c(h):= C_0 \pi_0 R_0(h) + C_1 \pi_1 R_1(h)\,,
\end{eqnarray}
or the following variant of \eqref{eqn:cost-sensitive}:
\begin{eqnarray}\label{eqn:cost-sensitive2}
\min_{h}R^{\bar{c}}(h):= C_0 R_0(h) + C_1 R_1(h)\,.
\end{eqnarray}
Then, the CS oracle classifier $h^{c*}$ under the cost-sensitive learning paradigm \eqref{eqn:cost-sensitive} can be calculated by $$h^{c*} (x) = \1\left(\frac{f_1(x)}{f_0(x)}> \frac{C_0}{C_1}\cdot\frac{\pi_0}{\pi_1}\right)\,,$$
and the CS oracle  $h^{\bar{c}*}$ under \eqref{eqn:cost-sensitive2}  can be calculated by $$h^{\bar{c}*} (x) = \1\left(\frac{f_1(x)}{f_0(x)}> \frac{C_0}{C_1}\right)\,.$$

Similar to its counterpart in the classical paradigm, the post-distortion CS oracle classifier is different from the pre-distortion CS oracle, and the pre-distortion CS oracle cannot be recovered in view of an unknown distortion scheme. Lemma \ref{prop: general cost-sensitive oracle under distortion} follows from arguments similar to the proof of Theorem 1 in the main text.
\begin{Lemma}\label{prop: general cost-sensitive oracle under distortion}
Suppose that $X|(Y=0)$ and $X|(Y=1)$ have probability density functions $f_0$ and $f_1$, and that class priors are $\pi_0$ and $\pi_1$ respectively.  Let $\beta_0$ and $\beta_1$ be the distortion rates of class $0$ and class $1$ respectively. Then, the oracle classifier under the cost-sensitive learning paradigm \eqref{eqn:cost-sensitive} regarding the post-distortion population is $$h^{c*}_{(\beta_0, \beta_1)}(x) = \1\left(\frac{f_1(x)}{f_0(x)}> \frac{1-\beta_0}{1-\beta_1}\cdot \frac{C_0}{C_1}\cdot\frac{\pi_0}{\pi_1}\right)\,.$$ Similarly, the oracle classifier under the paradigm \eqref{eqn:cost-sensitive2} regarding the post-distortion population is $$h^{\bar{c}*}_{(\beta_0, \beta_1)}(x) = \1\left(\frac{f_1(x)}{f_0(x)}> \frac{1-\beta_0}{1-\beta_1}\cdot \frac{C_0}{C_1}\right)\,.$$
\end{Lemma}
Lemma \ref{prop: general cost-sensitive oracle under distortion} implies that even if we have the entire post-distortion population, we can only mimic $h^{c*}_{(\beta_0, \beta_1)}$ or $h^{\bar{c}*}_{(\beta_0, \beta_1)}$. However, unless $\beta_0$ and $\beta_1$ are known or estimable, there is no hope to mimic $h^{c*}$ or $h^{\bar{c}*}$.

\section{Oracle classifiers when we relax the fixed class conditional densities assumption}\label{sec:: general assumption}

\begin{Proposition}\label{prop: most general classical oracle under distortion }
Suppose that pre-distortion, $X|(Y=0)$ and $X|(Y=1)$ have probability density functions $f_0$ and $f_1$, and that class priors are $\pi_0 = \p(Y=0)$ and $\pi_1 = \p(Y=1)$.  Let $\beta_0$ and $\beta_1$ be the distortion rates of class $0$ and class $1$ respectively. Further suppose that the post-distortion class conditional densities of features  are $f'_0$ and $f'_1$.  Then, the classical oracle classifier regarding the pre-distortion population is
\vspace{-0.05in}
$$h^*(x) = \1\left(\frac{f_1(x)}{f_0(x)} > \frac{\pi_0}{\pi_1}\right)\,,$$
\vspace{-0.05in}
and that regarding the post-distortion population is $$h^{*'}_{(\beta_0, \beta_1)}(x) = \1\left(\frac{f'_1(x)}{f'_0(x)} > \frac{1-\beta_0}{1-\beta_1}\cdot\frac{\pi_0}{\pi_1}\right)\,.
$$ %

\end{Proposition}
The proof is omitted due to its similarity to that for Theorem 1 in the main text.
Note that when $f'_1/ f'_0 = f_1 / f_0$, that is when the ratio of class conditional densities of features is preserved under data distortion,  the post-distortion classical oracle classifier $h^{*'}_{(\beta_0, \beta_1)}(x)$ reduces to $h^{*}_{(\beta_0, \beta_1)}(x)$ in Theorem 1, even if the class conditional densities themselves are changed. On the other hand, without assuming any relations between pre and post distortion feature distributions, $f_1/f_0$ cannot be recovered.

The invariance property (Theorem 2 in the main text) of Neyman-Pearson (NP) oracle classifiers no longer holds in general when the class conditional densities of features are different pre and post distortion. %
The next proposition illustrates sufficient and necessary conditions under which this invariance property does hold for a fixed $\alpha$. %

\begin{Proposition}\label{prop: general invariance}
Denote pre-distortion distributions of $X|(Y=0)$ and $X|(Y=1)$  by $f_0$ and $f_1$ and those post-distortion by $f_0'$ and $f_1'$. When $f'_1 / f'_0 = a\cdot  (f_1 / f_0)$  and
$$
a\cdot \min\{C\in\mathbb{R}: \p_{f_0}(f_1(X)/f_0(X) > C)\leq \alpha \}= \min \{C\in\mathbb{R}: \p_{f_0'}(f_1'(X)/f_0'(X) > C)\leq \alpha\}\,,
$$
for some $a>0$, the NP oracle classifier $\phi^*_{\alpha}$ defined in (4) in the main text is invariant under distortion at various rates $\beta_0$ (on class $0$) and $\beta_1$ (on class $1$), regardless of whether pre-distortion classes are balanced. Moreover, these conditions are also necessary for the invariance property.
\end{Proposition}

\begin{proof}
From the NP Lemma, it is easy to see that the two conditions are sufficient for the invariance property of the NP oracles. For the necessary part, again by the NP lemma, the NP oracles pre and post distortion can be written respectively as

$$
\phi_{\alpha}^*(x) = \1(f_1(x)/ f_0(x) > C_{\alpha}), \text{ and }\phi_{\alpha}^{*'}(x) = \1(f_1'(x)/ f_0'(x) > C_{\alpha}')\,,
$$
for some constants $C_{\alpha}$ and $C_{\alpha}'$ as determined in the NP Lemma. In other words,
$$
C_{\alpha} = \min\{C\in\mathbb{R}: \p_{f_0}(f_1(X)/f_0(X) > C)\leq \alpha \}\,,
$$
$$
C_{\alpha}' = \min\{C\in\mathbb{R}: \p_{f_0'}(f_1'(X)/f_0'(X) > C)\leq \alpha \}\,.
$$

Since $C_{\alpha}$ and $C_{\alpha}'$ are constants, to have $\phi_{\alpha}^*(x) = \phi_{\alpha}^{*'}(x)$, it is necessary to have $f'_1 / f'_0 = a\cdot  (f_1 / f_0)$ for some positive constants $a$, and this further demands $C_{\alpha}' = a\cdot C_{\alpha}$.
\end{proof}
Note that in general, the constant $a$ in Proposition \ref{prop: general invariance} depends on $\alpha$. In the following, we demonstrate that within certain distribution classes, the more general condition in Proposition \ref{prop: general invariance} falls back to the special case of unchanged class conditional feature distributions, while in others, there are $a\neq 1$ cases where class conditional feature distributions are different pre and post distortion.

{\bf Case I: Exponential Distribution}  Assume that $f_0(x)=\lambda_0e^{-\lambda_0x}$,  $f_1(x)=\lambda_1e^{-\lambda_1x}$; $f_0'(x)=\lambda_0'e^{-\lambda_0'x}$,  $f_1'(x)=\lambda_1'e^{-\lambda_1'x}$, where $x>0$. For identifiability concern, let us assume $\lambda_0<\lambda_1$, $\lambda_0'<\lambda_1'$.  Then, 
$$\frac{f_1(x)}{f_0(x)}=\frac{\lambda_1}{\lambda_0}e^{-(\lambda_1-\lambda_0)x}\,,$$
and 
$$\frac{f_1'(x)}{f_0'(x)}=\frac{\lambda_1'}{\lambda_0'}e^{-(\lambda_1'-\lambda_0')x}\,.$$
When we demand 
$$\frac{f_1'(x)}{f_0'(x)}=a\cdot\frac{f_1(x)}{f_0(x)}\ \ \ \ \ \ \forall  x\,,$$
it follows that 
\begin{equation}\label{ed1}
\lambda_1-\lambda_0=\lambda_1'-\lambda_0'\,,
\end{equation}
and 
\begin{equation}\label{ed2}
\frac{\lambda_1'}{\lambda_0'}=a\cdot\frac{\lambda_1}{\lambda_0}\,.
\end{equation}
Note that 
\begin{eqnarray*}
	P_{f_0}\left(\frac{f_1(X)}{f_0(X)}>C\right) &=&P_{f_0}\left(\frac{\lambda_1}{\lambda_0}e^{-(\lambda_1-\lambda_0)X}>C\right)\\
	&=&P_{f_0}\left(e^{-(\lambda_1-\lambda_0)X}>\frac{\lambda_0}{\lambda_1}C\right)\\
	&=&P_{f_0}\left(X<-\frac{1}{\lambda_1-\lambda_0}\ln \left(\frac{\lambda_0}{\lambda_1}C\right)\right)\\
	&=&1-\exp\left\{-\lambda_0\cdot\left[-\frac{1}{\lambda_1-\lambda_0}\ln \left(\frac{\lambda_0}{\lambda_1}C\right)\right]\right\}\\
	&=&1-\left(\frac{\lambda_0}{\lambda_1}C\right)^{\frac{\lambda_0}{\lambda_1-\lambda_0}}\,.
\end{eqnarray*}
To choose the minimum $C$ such that $P_{f_0}\left(\frac{f_1(X)}{f_0(X)}>C\right) \leq\alpha$, we get 
$$C_\alpha=\frac{\lambda_1}{\lambda_0}(1-\alpha)^{\frac{\lambda_1-\lambda_0}{\lambda_0}}\,.$$
Similarly, 
$$C_\alpha'=\frac{\lambda_1'}{\lambda_0'}(1-\alpha)^{\frac{\lambda_1'-\lambda_0'}{\lambda_0'}}\,.$$
Then  the condition $a\cdot C_\alpha=C_\alpha'$ implies that 
\begin{equation}\label{ed3}
a\cdot\frac{\lambda_1}{\lambda_0}(1-\alpha)^{\frac{\lambda_1-\lambda_0}{\lambda_0}}=\frac{\lambda_1'}{\lambda_0'}(1-\alpha)^{\frac{\lambda_1'-\lambda_0'}{\lambda_0'}}\,.
\end{equation}
For any given $0<\alpha<1$, combining three equations (\ref{ed1}), (\ref{ed2}) and (\ref{ed3}) implies that  
$$(1-\alpha)^{\frac{1}{\lambda_0}}=(1-\alpha)^{\frac{1}{\lambda_0'}}\,,$$
which implies that 
$\lambda_0=\lambda_0'$.  And then, $\lambda_1=\lambda_1'$ and $a=1$. Therefore, we have shown that when the class conditional feature distributions are restricted to the exponential distributions, the invariant property only occurs when $f_0 = f'_0$ and $f_1 = f'_1$.   

\

{\bf Case II: Gaussian Distribution}  Assume that $f_0: N(\mu_0,\sigma^2)$, $f_1: N(\mu_1,\sigma^2)$, $f_0': N(\mu_0',\sigma'^2)$,  and $f_1': N(\mu_1',\sigma'^2)$, where $\mu_0<\mu_1$, $\mu_0'<\mu_1'$, and $\sigma\neq\sigma'$. Then, 
$$\frac{f_1(x)}{f_0(x)}=\exp\left\{\frac{2(\mu_1-\mu_0)x+\mu_0^2-\mu_1^2}{2\sigma^2}\right\}$$
and 
$$\frac{f_1'(x)}{f_0'(x)}=\exp\left\{\frac{2(\mu_1'-\mu_0')x+\mu_0'^2-\mu_1'^2}{2\sigma'^2}\right\}\,.$$
To obtain 
$$\frac{f_1'(x)}{f_0'(x)}=a\cdot\frac{f_1(x)}{f_0(x)}\,,$$
the parameters $\mu_0,\mu_1,\sigma,\mu_0',\mu_1',\sigma', a$ must satisfy 
\begin{equation}\label{gd1}
\frac{2(\mu_1-\mu_0)}{2\sigma^2}=\frac{2(\mu_1'-\mu_0')}{2\sigma'^2}\,,
\end{equation}
and 
$$
a=\exp\left\{\frac{\mu_0'^2-\mu_1'^2}{2\sigma'^2}-\frac{\mu_0^2-\mu_1^2}{2\sigma^2}\right\}\,.
$$
Furthermore, denote by $\Phi(\cdot)$ the cumulative distribution function of standard normal distribution, $C_\alpha=\min_C\left\{C\in R:P_{f_0}\left(\frac{f_1(X)}{f_0(X)}>C\right) \leq\alpha\right\}$ and $C_\alpha'=\min_C\left\{C\in R:P_{f_0'}\left(\frac{f_1'(X)}{f_0'(X)}>C\right) \leq\alpha\right\}$. 
\begin{eqnarray*}
	P_{f_0}\left(\frac{f_1(X)}{f_0(X)}>C\right) &=&P_{f_0}\left(\exp\left\{\frac{2(\mu_1-\mu_0)X+\mu_0^2-\mu_1^2}{2\sigma^2}\right\}>C\right)\\
	&=&P_{f_0}\left(2(\mu_1-\mu_0)X+\mu_0^2-\mu_1^2>2\sigma^2\ln C\right)\\
	&=&P_{f_0}\left(X>\frac{2\sigma^2\ln C+\mu_1^2-\mu_0^2}{2(\mu_1-\mu_0)}\right)\\
	&=&P_{f_0}\left(\frac{X-\mu_0}{\sigma}>\frac{2\sigma^2\ln C+(\mu_1-\mu_0)^2}{2(\mu_1-\mu_0)\sigma}\right)\,.
\end{eqnarray*}
Based on $P_{f_0}\left(\frac{f_1(X)}{f_0(X)}>C\right) \leq\alpha$, we get
$$\Phi^{-1}(1-\alpha)\leq \frac{2\sigma^2\ln C+(\mu_1-\mu_0)^2}{2(\mu_1-\mu_0)\sigma}\,,$$
where $\Phi^{-1}(\cdot)$ is the inverse function of $\Phi(\cdot)$, that is, 
$$C\geq \exp\left\{\frac{2\sigma(\mu_1-\mu_0)\Phi^{-1}(1-\alpha)-(\mu_1-\mu_0)^2}{2\sigma^2}\right\}\,.$$
Therefore, 
$$C_\alpha=\exp\left\{\frac{2\sigma(\mu_1-\mu_0)\Phi^{-1}(1-\alpha)-(\mu_1-\mu_0)^2}{2\sigma^2}\right\}\,.$$
Similarly, 
$$C_\alpha'=\exp\left\{\frac{2\sigma'(\mu_1'-\mu_0')\Phi^{-1}(1-\alpha)-(\mu_1'-\mu_0')^2}{2\sigma'^2}\right\}.$$
From the relationship $a\cdot C_\alpha=C_\alpha'$, we can obtain 
\begin{eqnarray}\label{gd2}
&&\frac{\mu_0'^2-\mu_1'^2}{2\sigma'^2}-\frac{\mu_0^2-\mu_1^2}{2\sigma^2}+\frac{2\sigma(\mu_1-\mu_0)\Phi^{-1}(1-\alpha)-(\mu_1-\mu_0)^2}{2\sigma^2}\nonumber\\
&=&\frac{2\sigma'(\mu_1'-\mu_0')\Phi^{-1}(1-\alpha)-(\mu_1'-\mu_0')^2}{2\sigma'^2}\,,
\end{eqnarray}
i.e., 
\begin{eqnarray*}
&&\frac{\mu_0'^2-\mu_1'^2}{2\sigma'^2}+\frac{(\mu_1'-\mu_0')^2}{2\sigma'^2}-\frac{\mu_0^2-\mu_1^2}{2\sigma^2}-\frac{(\mu_1-\mu_0)^2}{2\sigma^2}\\
&=&\frac{(\mu_1'-\mu_0')\Phi^{-1}(1-\alpha)}{\sigma'}-\frac{(\mu_1-\mu_0)\Phi^{-1}(1-\alpha)}{\sigma}\,,
\end{eqnarray*}
which is equivalent to, 
\begin{equation}\label{gd3}
\frac{\mu_0'(\mu_0'-\mu_1')}{\sigma'^2}-\frac{\mu_0(\mu_0-\mu_1)}{\sigma^2}=\left[\frac{(\mu_1'-\mu_0')}{\sigma'}-\frac{(\mu_1-\mu_0)}{\sigma}\right]\Phi^{-1}(1-\alpha)\,.
\end{equation}
From equation (\ref{gd1})
\begin{equation}\label{gd4}
\frac{\mu_1'-\mu_0'}{\sigma'}=\frac{\sigma'}{\sigma^2}(\mu_1-\mu_0)\,.
\end{equation}
Putting (\ref{gd4}) into (\ref{gd3}), 
$$\frac{(\mu_0-\mu_1)}{\sigma^2}(\mu_0'-\mu_0)=\left[\frac{\sigma'}{\sigma^2}(\mu_1-\mu_0)-\frac{(\mu_1-\mu_0)}{\sigma}\right]\Phi^{-1}(1-\alpha)\,,$$
that is, 
$$\Phi^{-1}(1-\alpha)=\frac{\mu_0-\mu_0'}{\sigma'-\sigma}\,.$$
Putting the above arguments together, we have shown that under Gaussian distributions, for a given $\alpha \in (0,1)$,  the invariance property is satisfied precisely when 

$$\frac{(\mu_1-\mu_0)}{\sigma^2}=\frac{(\mu_1'-\mu_0')}{\sigma'^2}\,,$$
$$\Phi^{-1}(1-\alpha)=\frac{\mu_0-\mu_0'}{\sigma'-\sigma}\,,$$ and 
$$
a=\exp\left\{\frac{\mu_0'^2-\mu_1'^2}{2\sigma'^2}-\frac{\mu_0^2-\mu_1^2}{2\sigma^2}\right\}\,.
$$

\

\ 

{\bf Example of Case II:}
Let $f_0: N(0,2^2)$, $f_1: N(1,2^2)$ and $f_0': N(-4,4^2)$, and $f_1': N(0,4^2)$. We show that when $\alpha = 0.023$, the invariant property holds. First, it is easy to check that the above three equations hold with these density specifications and the choice of $\alpha$. In the following, we provide an alternative direct proof.

Note that   
$$\frac{f_1(x)}{f_0(x)}=\exp\left\{\frac{2x-1}{8}\right\}\,,$$
and 
$$\frac{f_1'(x)}{f_0'(x)}=\exp\left\{\frac{8x+16}{32}\right\}\,,$$
Hence, 
$$\frac{f_1'(x)}{f_0'(x)}=\exp\left\{\frac{5}{8}\right\}\cdot\frac{f_1(x)}{f_0(x)}\,.$$
We can take $a=\exp\left\{\frac{5}{8}\right\}$. Let $\alpha=0.023$. Then  $\Phi^{-1}(1-\alpha)=2$. We solve for $C_{\alpha}$ and $C'_{\alpha}$ from 
 $$P_{f_0}\left(\frac{f_1(X)}{f_0(X)}>C_\alpha\right)=\alpha\ \ \ \text{and}\ \ \ \ P_{f_0'}\left(\frac{f_1'(X)}{f_0'(X)}>C_\alpha'\right)=\alpha\,.$$
That is,  
 $$P_{f_0}\left(\exp\left\{\frac{2X-1}{8}\right\}>C_\alpha\right)=\alpha\ \ \ \text{and}\ \ \  P_{f_0'}\left(\exp\left\{\frac{8X+16}{32}\right\}>C_\alpha'\right)\,,$$
Or equivalently,
  $$P_{f_0}\left(X>\frac{8\ln C_\alpha+1}{2}\right)=\alpha\ \ \ \text{and}\ \ \  P_{f_0'}\left(X>4\ln C_\alpha'-2\right) = \alpha \,.$$
That is, 
 $$\frac{(8\ln C_\alpha+1)/2}{2}=2\ \ \ \text{and}\ \ \   \frac{4\ln C_\alpha'-2-(-4)}{4}=2,$$
 which implies that
 $$C_\alpha=\exp\left\{\frac{7}{8}\right\}\ \ \ \text{and}\ \ \  C_\alpha'=\exp\left\{\frac{3}{2}\right\}\,.$$
 Obviously, 
 $$a\cdot C_\alpha=C_\alpha'\,,$$
i.e., 
$$a\cdot\min_C\left\{C\in R:P_{f_0}\left(\frac{f_1}{f_0}>C\right) \leq\alpha\right\}=\min_C\left\{C\in R: P_{f_0'}\left(\frac{f_1'}{f_0'}>C\right) \leq\alpha\right\}\,.$$
Therefore, we have constructed a concrete NP oracle invariant example in which $f_0 \neq f_0'$ and $f_1 \neq f'_1$.

\section{Sparsity-inducing methods in selecting meaningful topics}\label{NP-sLDA}

\begin{figure}
\includegraphics[scale=0.6]{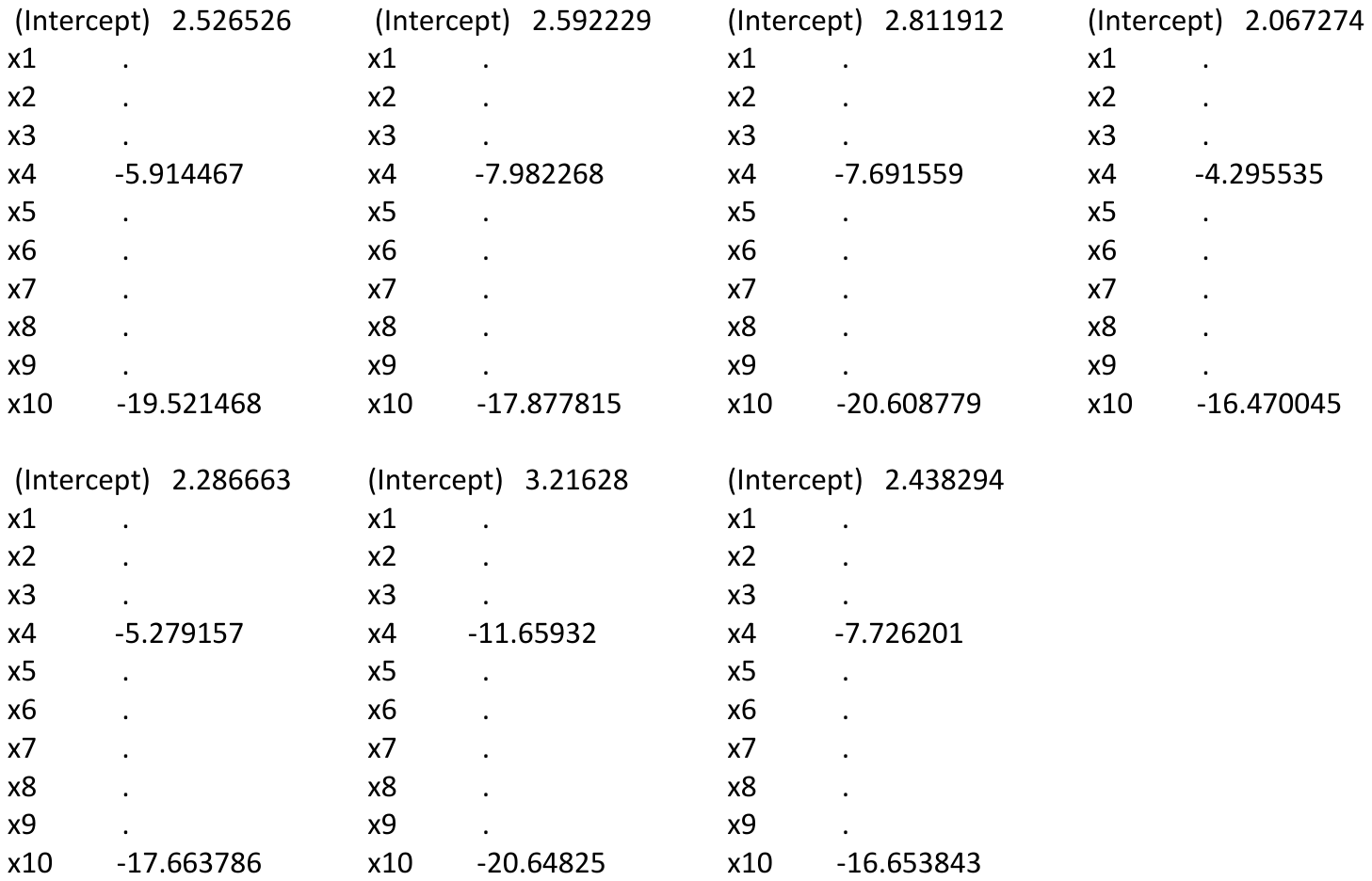}
\caption{regression coefficients for the $7$ splits in NP-sLDA, repetition $1$.} \label{fig:rep1}
\end{figure}

\begin{figure}
\includegraphics[scale=0.6]{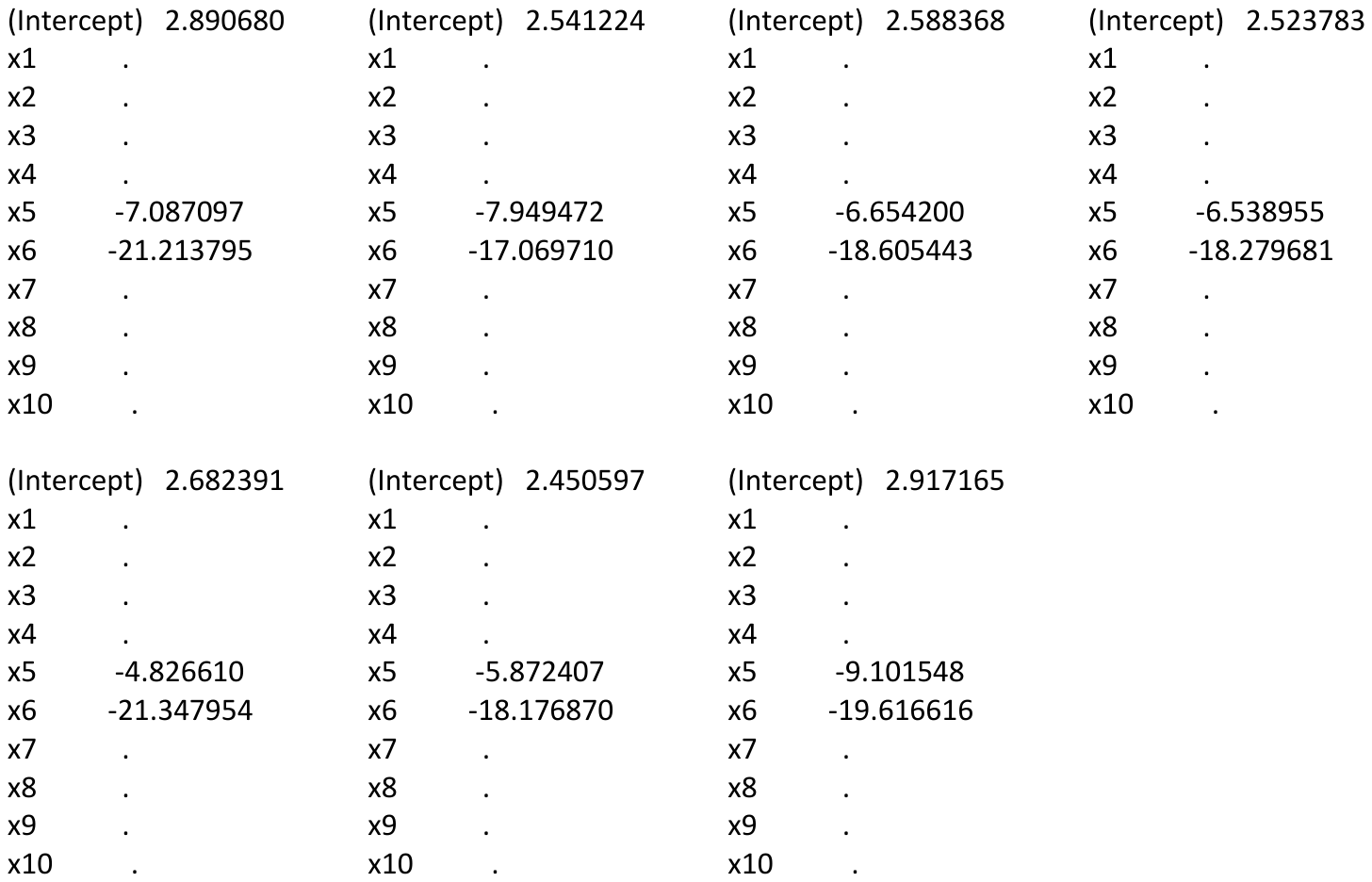}
\caption{regression coefficients for the $7$ splits in NP-sLDA, repetition $2$.} \label{fig:rep2}
\end{figure}

Among the implemented methods, NP-sLDA performs the best in terms of power and it is a penalized sparsity-inducing method, which means it eliminates certain unimportant features as part of the classifier training process.  In this section, we elaborate that such methods are effective in terms of selecting meaningful topics. In particular, we look at results from the first two random repetitions under {\sf Setting 1} in Section 4.5.2 (random seed being set and results are readily available online) with $K=10$. In the first repetition, Table \ref{table:rep1} displays the selected ten topics and it's obvious that only topics $4$ and $10$ are the strike-related topics. Following the common practice of NP umbrella algorithms, we randomly split the training data $M$ times for training the scoring function and thresholds. Here we use $M = 7$, and the final classifier is a majority vote.  Figure \ref{fig:rep1} shows that, over the seven splits, NP-sLDA consistently selects only topics $4$ and $10$, and all the rest of the topics have corresponding coefficient $0$. Similarly, in repetition 2, Table \ref{table:rep2} shows that only topics $5$ and $6$ are the strike-related topics, and Figure \ref{fig:rep2} shows that  NP-sLDA consistently selects topics $5$ and $6$ over the $7$ splits. In summary, these sparsity-inducing methods, such as NP-sLDA, help select meaningful topics. 

\begin{CJK*}{UTF8}{gbsn}
\begin{table*}
		\fontsize{6pt}{9pt}\selectfont
		\begin{center}
			\setlength\tabcolsep{2pt}
			\begin{tabular}{ccccccccccc}
				\hline                  
				topic 1&罢工 &终于 &学校 &时间 &一下&事件&彻底 &哼哼&开&对\cr
				&strike&finally&school&time&a bit&event&complete&humph&open&right\cr
				&电话&集体&分钟& 失望 &胃&为了&好多 &疑问&多少&忙\cr
				&phone&collective&minute&disappoint&stomach&for&many&question&how many&busy\cr
				\hline
				topic 2&罢工&今天&上班&发生 &年 &上 &发现& 问题 &种 &衰\cr
				&strike&today&work&happen&year&go&discover&problem&type&decline\cr
				&回来 &太阳 &话& 公交车 &宿舍& 冷 &块& 过节& 东西 &思考\cr
				&come back&Sun&words&bus&dorm&cold&block&festival&things&think\cr
				\hline                 
				topic 3&人 &让& 说 &吃& 时候& 事 &罢课& 过 &哈哈 &小\cr
				&people&let&speak&eat&time&thing&student strike&pass&haha&small\cr
				&里 &妈妈& 老师& 今晚& 去 &很多 &找 &出门& 最近& 班\cr
				&inside&mom&teacher&tonight&go&many&find&go out&recent&class\cr
				\hline                  
				topic 4& 年 &公司& 员工& 中& 工人& 工作& 工资 &最后& 月 &鄙视\cr
				&year&company&employee&within&worker&work&salary&finally&month&despise\cr
				&小时& 后& 广州& 抗议& 今日& 知道& 请 &月日& 要求 &中国\cr
				&hour&after&Guangzhou&protest&today&know&please&month-date&request&China\cr
				\hline                
				topic 5& 抓& 狂& 罢工 &电脑 &泪& 早上 &现在 &抓狂 &天气 &回家\cr
				&clutch&crazy&strike&computer&tear&morning&now&go crazy&weather&go home\cr
				&想 &潮湿& 下午& 结果& 集 &继续& 部 &修 &人& 委屈\cr
				&think&moist&afternoon&result&gather&continue&department&fix&people&be wronged\cr
				\hline                  
				topic 6& 罢工 &去 &做& 次& 能& 生病 &地 &系& 偷笑 &没有\cr
				&strike&go&do&times&can&sick&ground&systems&smirk&without\cr
				&睡觉 &鼻屎& 挖 &第一& 今晚& 怒 &回 &真的& 叫 &汗\cr
				&sleep&mucus&pick&first&tonight&angry&back&real&shout&sweat\cr
				\hline               
				topic 7& 天 &手机 &还是& 知道& 能& 竟然& 突然 &说& 玩& 这个\cr
				&day&cellphone&still&know&can&unexpectedly&suddenly&say&play&this\cr
				&出来 &换 &已经& 点 &郁闷 &鼓掌 &听 &一下& 真是& 好不容易\cr
				&out&exchange&already&bit&depressed&applaud&listen&a bit&really&hard\cr
				\hline                  
				topic 8& 罢工& 想 &可怜& 居然& 买 &发& 明天& 累 &点 &但是\cr
				&strike&think&pity&unexpectedly&buy&give&tomorrow&tired&bit&but\cr
				&星期& 然后 &休息 &家里 &半 &悲伤 &一直& 本来& 听说& 心情\cr
				&week&therefore&rest&home&half&sad&always&originally&heard&mood\cr
				\hline              
				topic 9&罢工 &草草& 明天 &可以 &开始& 好好 &真的 &新闻 &爱& 开\cr
				&strike&hastily&tomorrow&can&start&nicely&really&news&love&open\cr
				&心情 &点 &还有& 刚刚& 这个& 之后 &一定& 为什么& 晚& 上午\cr
				&mood&a bit&also&just&this&after&must&why&evening&morning\cr
				\hline                
				topic 10& 的士 &汕头& 出租车& 司机& 现在& 车& 罢工 &打& 下& 辆\cr
				&taxi&Shantou&taxi&driver&now&car&strike&call&get off&vehicle\cr
				&营运 &三& 原因& 政府& 集体& 今日& 希望& 四 &市民& 月日\cr
				&operate&three&reason&government&collective&today&hope&four&citizen&month-date\cr
				\hline
			\end{tabular}
			\caption{top $20$ keywords for the ten topics selected from repetition $1$.}
			\label{table:rep1}
		\end{center}
	\end{table*}

\begin{table*}
		\fontsize{6pt}{9pt}\selectfont
		\begin{center}
			\setlength\tabcolsep{2pt}
			\begin{tabular}{ccccccccccc}
				\hline                  
				topic 1&今天 &天 &可以& 没有& 开始& 点& 真的 &日子& 明天& 能\cr
				&today&day&can&without&start&bit&really&day&tomorrow&can\cr
				&前 &回家& 还有 &吃饭& 吃 &号 &那些 &地铁& 哈哈& 玩\cr
				&forward&go home&also&eat&eat&day&those&subway&haha&play\cr
				\hline                   
				topic 2&罢工 &上& 上班& 能 &终于& 抓狂 &拿 &小时 &里 &东西\cr
				&strike&go to&work&can&finally&go crazy&get&hour&inside&thing\cr
				&真是& 三& 为了& 生活& 之后 &超级& 只是 &开心 &觉得& 对\cr
				&really&three&for&life&after&super&just&happy&feel&right\cr
				\hline                 
				topic 3&罢工 &去& 系& 做& 今日 &地 &睡觉 &后& 起来& 听\cr
				&strike&go&be&do&today&ground&sleep&after&get up&listen\cr
				&搞& 过 &怒 &公交& 求& 人& 甘 &吃 &街 &说\cr
				&do&over&angry&public transportation&beg&people&willing&eat&street&speak\cr
				\hline                  
				topic 4& 想 &人 &让& 说& 罢课& 累 &发& 衰& 生病 &找\cr
				&think&people&let&speak&student strike&tired&happen&decline&sick&find\cr
				&次 &鄙视 &很多& 新& 但是& 哦& 感冒 &虽然& 委屈 &竟然\cr
				&time&despise&many&new&but&oh&a cold&although&be wronged&unexpectedly\cr
				\hline                
				topic 5& 年& 公司 &月 &员工 &工人& 工资& 最后& 月日& 工作 &还是\cr
				&year&company&month&employee&worker&salary&finally&month-date&work&still\cr
				&集体& 第一& 国际 &次 &要求 &劳动& 无法& 机场& 买 &法国\cr
				&collective&first&international&time&demand&labor&unable&airport&buy&France\cr
				\hline                  
				topic 6& 罢工& 的士& 汕头& 现在& 出租车 &司机& 车& 打 &集体& 辆\cr
				&strike&taxi&Shantou&now&taxi&driver&car&call&collective&vehicle\cr
				&下& 广州 &出门& 已经& 政府 &事件& 出 &路& 钱& 问题\cr
				&get off&Guangzhou&go out&already&government&event&out&street&money&problem\cr
				\hline               
				topic 7& 可怜& 小& 结果 &偷笑& 发现& 昨天 &今晚 &早上& 能& 一直\cr
				&pity&small&result&smirk&find&yesterday&tonight&morning&can&always\cr
				&生病 &竟然 &郁闷& 开 &星期& 罢工& 三 &今天 &出来 &结局\cr
				&sick&unexpectedly&depressed&open&week&strike&three&today&go out&end\cr
				\hline                  
				topic 8& 罢工 &天& 知道& 天气 &时候& 挖 &鼻屎& 太阳 &种 &周\cr
				&strike&day&know&weather&time&pick&mucus&Sun&type&week\cr
				&今天& 时间 &电视 &突然 &奥特曼 &好像 &应该& 全部& 水 &点\cr
				&today&time&TV&sudden&Ultraman&maybe&should&whole&water&bit\cr
				\hline              
				topic 9&罢工 &抓 &狂& 泪& 电脑& 居然 &今晚& 鼓掌& 泪泪& 学校\cr
				&strike&clutch&crazy&tear&computer&unexpectedly&tonight&applaud&tear&school\cr
				&事 &搞到& 部 &学生& 结 &啊啊 &手机& 明天& 闹 &闹钟\cr
				&thing&get&department&student&form&ah&cellphone&tomorrow&alarm&alarm clock\cr
				\hline                
				topic 10& 罢工 &手机& 中 &最近 &哼哼& 一下& 哈哈& 电梯 &停播 &玩\cr
				&strike&cellphone&within&recent&humph&a bit&haha&elevator&stop playing&play\cr
				&女& 怒骂& 分钟& 时候 &过 &晚& 深圳 &第一& 迟到& 下班\cr
				&female&curse&minute&time&over&late&Shenzhen&first&late&off work\cr
				\hline
			\end{tabular}
			\caption{top $20$ keywords for the ten topics selected from repetition $2$.}
			\label{table:rep2}
		\end{center}
	\end{table*}
\end{CJK*}

\section{Proof and generalization of proposition 1 in main text}\label{proof:prop1}

Proposition 1 in the main text follows as a special case of the next Proposition.  %
Proposition \ref{prop:A1} below explores the relationship between type I error $R_0(\cdot)$, the distortion rate $\beta_0$ of class $0$ and the class size ratio $\pi_0/\pi_1$ for the classical post-distortion oracle classifier $h^*_{\beta_0, \pi_0}$. %

\begin{Proposition}\label{prop:A1}
Suppose probability densities of class $0$ ($X|Y=0$) and  class $1$ ($X|Y=1$) follow distributions $\mathcal{N}(\mu_0,\Sigma)$ and $\mathcal{N}(\mu_1,\Sigma)$ respectively; class $0$ composes $\pi_0\in (0, 1)$ proportion of the population and $\beta_0\in(0,1)$ is the censorship rate of class $0$ (i.e., the proportion of class $0$ posts that were removed from some government censorship scheme). Suppose class 1 is not distorted (i.e., $\beta_1 = 0$). Let  $h^*_{\beta_0, \pi_0}$ be the classical oracle classifier in the post-distortion population.   Then the type I error of $h^*_{\beta_0, \pi_0}$ (regarding either the pre-distortion or the post-distortion population) is calculated as :
\begin{equation}\label{type_i_error_censored}
R_0(h^*_{\beta_0, \pi_0}) = \Phi\left(\frac{-\frac{1}{2}C-\log\left((1-\beta_0)p\right)}{\sqrt{C}}\right)\,,
\end{equation}
where $C = (\mu_0 - \mu_1)^{\top}\Sigma^{-1}(\mu_0 - \mu_1)$ and $p = \pi_0 / (1 - \pi_0)$.  Equation \eqref{type_i_error_censored} implies that
\begin{enumerate}
\item Keeping $\pi_0$ fixed (hence $p$ is fixed), $R_0(h^*_{\beta_0, \pi_0})$ is a monotone increasing function of the class $0$ censorship rate $\beta_0\in (0, 1)$. Moreover, we have i). if $pe^{3C/2} \leq  1$,  $R_0(h^*_{\beta_0, \pi_0})$ is a concave function of $\beta_0\in(0, 1)$; and ii). if $pe^{3C/2} > 1$, $R_0(h^*_{\beta_0, \pi_0})$ is a convex function of $\beta_0$ for $\beta_0 \in \left(0, 1 - \frac{1}{pe^{3C/2}}\right)$, and a concave function for $\beta_0 \in \left(1 - \frac{1}{pe^{3C/2}}, 1\right)$.
\item Keeping $\beta_0$ fixed, $R_0(h^*_{\beta_0, \pi_0})$ is a monotone decreasing function of the class ratio $ p = \pi_0/(1-\pi_0)$. In other words, the larger the proportion of class $0$ in the uncensored population, the smaller the type I error of $h^*_{\beta_0, \pi_0}$. Moreover, $R_0(h^*_{\beta_0, \pi_0})$ is a convex function of $p$ for $p > \frac{1}{(1-\beta_0)e^{3C/2}}$, and it is a concave function of $p$ for $p \leq \frac{1}{(1-\beta_0)e^{3C/2}}$.
\end{enumerate}
\end{Proposition}
\begin{proof}
Since equation (2) in the main text is  the decision boundary of $h^*_{\beta_0, \pi_0}$, we have
\begin{equation*}
R_0(h^*_{\beta_0, \pi_0})  = P_{X\sim\mathcal{N}(\mu_0,\Sigma)}\left\{X^{\top} \Sigma^{-1} (\mu_0-\mu_1) -\frac{1}{2} (\mu_0-\mu_1)^{\top}\Sigma^{-1}(\mu_0+\mu_1) + \log\left(\frac{(1-\beta_0) \pi_0}{\pi_1}\right)\leq 0\right\}\,.
\end{equation*}
For $X$ in class $0$, $X^{\top} \Sigma^{-1} (\mu_0-\mu_1)\eqdef Z'\sim\mathcal{N}(\mu^{\top}_0\Sigma^{-1}(\mu_0-\mu_1),(\mu_0-\mu_1)^{\top}\Sigma^{-1}(\mu_0-\mu_1))$. Therefore,
\small{
\begin{align*}
R_0(h^*_{\beta_0, \pi_0})  &= P_{Z'\sim\mathcal{N}(\mu^{\top}_0\Sigma^{-1}(\mu_0-\mu_1),(\mu_0-\mu_1)^{\top}\Sigma^{-1}(\mu_0-\mu_1))}\left\{Z'\leq \frac{1}{2}(\mu_0-\mu_1)^{\top}\Sigma^{-1}(\mu_0+\mu_1)-\log\left(\frac{(1-\beta_0)\pi_0}{\pi_1}\right)\right\}\cr
&= \Phi\left(\frac{-\frac{1}{2}(\mu_0-\mu_1)^{\top}\Sigma^{-1}(\mu_0-\mu_1)-\log\left(\frac{(1-\beta_0)\pi_0}{\pi_1}\right)}{\sqrt{(\mu_0-\mu_1)^{\top}\Sigma^{-1}(\mu_0-\mu_1)}}\right)\,.
\end{align*}
}
Regarding part 1, for fixed $\pi_0$,  let $f(\beta_0) = R_0(h^*_{\beta_0, \pi_0})$.
$$
f'(\beta_0) = \phi\left(\frac{-\frac{1}{2}C-\log\left((1-\beta_0)p\right)}{\sqrt{C}}\right)\cdot \frac{1}{\sqrt{C}(1-\beta_0)}\,,
$$
where $\phi(\cdot)$ is the probability density function of the standard normal random variable.    This implies that for $\beta_0\in (0, 1)$, $f'(\cdot)$ is positive, so $R_0(h^*_{\beta_0, \pi_0})$ is a monotone increasing function of $\beta_0$ for fixed $\pi_0$. Taking the second derivative of $f$, we have
$$
f''(\beta_0) = \phi'\left(\frac{-\frac{1}{2}C-\log\left((1-\beta_0)p\right)}{\sqrt{C}}\right)\cdot \frac{1}{C(1-\beta_0)^2} + \phi\left(\frac{-\frac{1}{2}C-\log\left((1-\beta_0)p\right)}{\sqrt{C}}\right)\cdot \frac{1}{\sqrt{C}(1-\beta_0)^2}\,.
$$
Let $g(w) = \phi'(w) + \sqrt{C} \phi(w)$.  Then

$$
g(w) = \frac{1}{\sqrt{2\pi}}e^{-\frac{w^2}{2}}\cdot(-w) + \frac{\sqrt{C}}{\sqrt{2\pi}}e^{-\frac{w^2}{2}}\,.
$$
Note that $g(w)>0$ iff $w < \sqrt{C}$.

Therefore, $f''(\beta_0) > 0$ iff $g(\frac{-\frac{1}{2}C-\log\left((1-\beta_0)p\right)}{\sqrt{C}}) > 0$ iff $\frac{-\frac{1}{2}C-\log\left((1-\beta_0)p\right)}{\sqrt{C}}    < \sqrt{C}$ iff $\beta_0 < 1 - \frac{1}{p e^{3C/2}}$.  Similarly $f''(\beta_0) < 0$ iff $\beta_0 > 1 - \frac{1}{pe^{3C/2}}$.

Regarding part 2, for fixed $\beta_0$, let $k(p) = R_0(h^*_{\beta_0, \pi_0})$, then

$$
k'(p) = \phi\left(\frac{-\frac{1}{2}C-\log\left((1-\beta_0)p\right)}{\sqrt{C}}\right)\cdot\frac{-1}{\sqrt{C}p}\,.
$$
Clearly, $k'(p) < 0$ for all $p >0$.
$$
k''(p) = \phi'\left(\frac{-\frac{1}{2}C-\log\left((1-\beta_0)p\right)}{\sqrt{C}}\right)\cdot \frac{1}{Cp^2} + \phi\left(\frac{-\frac{1}{2}C-\log\left((1-\beta_0)p\right)}{\sqrt{C}}\right)\cdot \frac{1}{\sqrt{C}p^2}\,.
$$
Note that $k''(p)> 0$ iff $\frac{-\frac{1}{2}C-\log\left((1-\beta_0)p\right)}{\sqrt{C}}    < \sqrt{C}$ iff $p > \frac{1}{(1-\beta_0)e^{3C/2}}$.
\end{proof}
The constant $C$ can be considered as a measure of separability of the two classes.
Note that when $p = 1$, that is when $\pi_0 = 1- \pi_0 = 1/2$, if $C$ is large (i.e., it is easy to separate the two classes), $1/(pe^{3C/2})\approx 0$, then $R_0(h^*_{\beta_0, \pi_0})$ is a convex function of $\beta_0 \in (0, 1)$. On the other hand, when $C$ is so small (i.e., two classes are hard to separate) that $p e^{3C/2}\leq 1$, $R_0(h^*_{\beta_0, \pi_0})$ is a concave function of $\beta_0 \in (0, 1)$.

\section{Neyman-Pearson Lemma}\label{sec:Neyman-Pearson Lemma}
The oracle classifier under the NP paradigm (NP oracle) arises from its close connection to the Neyman-Pearson Lemma in statistical hypothesis testing.
Hypothesis testing bears strong resemblance to binary classification if we assume the following  model. Let $P_1$ and $P_0$ be two \textit{known} probability distributions on $\mathcal{X}\subset \mathbb{R}^d$.
Assume that $Y\sim \text{Bern}(\zeta)$ for some $\zeta \in (0,1)$, and the conditional distribution of $X$ given $Y$ is $P_Y$.
Given such a model, the goal of statistical hypothesis testing is to determine if we should reject the null hypothesis that $X$ was generated from $P_0$.
To this end, we construct a randomized test $\phi:\mathcal{X} \to [0,1]$ that rejects the null with probability $\phi(X)$.
Two types of errors arise: type~I error occurs when $P_0$ is rejected yet $X\sim P_0$, and type~II error occurs when $P_0$ is not rejected yet $X\sim P_1$.
The Neyman-Pearson paradigm in hypothesis testing amounts to choosing $\phi$ that solves the following constrained optimization problem
$$
\text{maximize } \E[\phi(X)|Y=1]\,,
\text{ subject to }  \E[\phi(X)|Y= 0 ]\leq\alpha\,,
$$
where $\alpha \in (0,1)$ is the significance level of the test. A solution to this constrained optimization problem is called  \emph{a most powerful test} of level $\alpha$. The Neyman-Pearson Lemma gives mild sufficient conditions for the existence of such a test.

\begin{Lemma}[Neyman-Pearson Lemma]\label{lemma:NP}
Let $P_1$ and $P_0$ be two probability measures with densities $f_1
$ and $f_0$ respectively, and denote the density ratio as $r(x)=f_1(x)/f_0(x)$.
For a given significance level $\alpha$, let $C_{\alpha}$ be such that
$P_0\{r(X)>C_{\alpha}\}\leq\alpha$ and $P_0\{r(X)\geq C_{\alpha}\}\geq\alpha$.
Then,
the most powerful test of level $\alpha$ is
\begin{equation*}
\phi^*_{\alpha}(X)=\left\{
 \begin{array}{ll}
     1 & \text{if $\,\,r(X)>C_{\alpha}$}\,,\\
     0 & \text{if $\,\,r(X)<C_{\alpha}$}\,,\\
     \frac{\alpha-P_0\{r(X)>C_{\alpha}\}}{P_0\{r(X)=C_{\alpha}\}} & \text{if $\,\,r(X)=C_{\alpha}$}\,.
   \end{array}      \right.
\end{equation*}
\end{Lemma}
Under mild continuity assumption, we take the \emph{NP oracle classifier}
\begin{align}\label{eq::oracle}
\phi^*_{\alpha}(x) \,=\, \1\{f_1(x)/f_0(x) > C_{\alpha}\} \,=\, \1\{r(x) > C_{\alpha}\}\,,
\end{align}
as our plug-in target for NP classification. %

\end{document}